\documentclass[a4paper,fleqn]{cas-sc}

\usepackage[numbers]{natbib}

\usepackage{pifont}
\usepackage{twemojis}
\usepackage{enumitem}

\usepackage{pgfplots}
\pgfplotsset{compat=1.9}
\usepgfplotslibrary{statistics}
\usetikzlibrary{trees}

\def\twodigits#1{\ifnum#1<10 0\fi 
\number#1}

\def\tsc#1{\csdef{#1}{\textsc{\lowercase{#1}}\xspace}}
\tsc{WGM}
\tsc{QE}
\tsc{EP}
\tsc{PMS}
\tsc{BEC}
\tsc{DE}

\begin{document}
\let\WriteBookmarks\relax
\def\floatpagepagefraction{1}
\def\textpagefraction{.001}

\shorttitle{Enhancing Program Comprehension through Impermanent Identifiers}

\shortauthors{Guerra, Eduardo et~al.}

\title [mode = title]{Impermanent Identifiers: Enhanced Source Code Comprehension and Refactoring}

\author[1]{Eduardo Martins Guerra}[auid=000,bioid=1,
                        orcid=0000-0001-5555-3487]

\cormark[1]

\fnmark[1]

\ead{eduardo.guerra@unibz.it}

\ead[url]{https://www.unibz.it/en/faculties/engineering/academic-staff/person/43879-eduardo-martins-guerra}

\affiliation[1]{organization={Free University of Bozen-Bolzano},
city={Bozen-Bolzano},
country={Italy}}

\author[2]{André A. S. Ivo}[orcid=0000-0001-6192-7705]
\ead{andre.ivo@inpe.gov.br}

\author[2]{Fernando O. Pereira}[orcid=0009-0006-2360-1281]
\ead{fernando.opc@gmail.com}

\author[1]{Romain Robbes}[orcid=0000-0003-4569-6868]
\ead{romain.robbes@gmail.com}

\author[3]{Andrea Janes}[orcid=0000-0002-1423-6773]
\ead{andrea.janes@unibz.it}

\author[4]{Fábio Fagundes Silveira}[orcid=0000-0002-2063-2959
]
\ead{fsilveira@unifesp.br}

\affiliation[2]{organization={National Institute for Space Research (INPE)},
city={São José dos Campos},
state={São Paulo},
    country={Brazil}}

\affiliation[4]{organization={Federal University of São Paulo},
city={São José dos Campos},
state={São Paulo},
    country={Brazil}}

\affiliation[3]{organization={Free University of Bozen-Bolzano},
city={Bolzano},
country={Italy}}

\cortext[cor1]{Corresponding author}

\newcommand{\revone}[1]{\textcolor{red}{#1}}
\newcommand{\revtwo}[1]{\textcolor{orange}{#1}}
\newcommand{\revthree}[1]{\textcolor{green}{#1}}

\begin{abstract}
In response to the prevailing challenges in contemporary software development, this article introduces an innovative approach to code augmentation centered around \textit{Impermanent Identifiers}. The primary goal is to enhance the software development experience by introducing dynamic identifiers that adapt to changing contexts, facilitating more efficient interactions between developers and source code, ultimately advancing comprehension, maintenance, and collaboration in software development.
Additionally, this study rigorously evaluates the adoption and acceptance of \textit{Impermanent Identifiers} within the software development landscape. Through a comprehensive empirical examination, we investigate how developers perceive and integrate this approach into their daily programming practices, exploring perceived benefits, potential barriers, and factors influencing its adoption.
In summary, this article charts a new course for code augmentation, proposing \textit{Impermanent Identifiers} as its cornerstone while assessing their feasibility and acceptance among developers. This interdisciplinary research seeks to contribute to the continuous improvement of software development practices and the progress of code augmentation technology.
\end{abstract}

\begin{keywords}
source code identifiers \sep program comprehension \sep software refactoring \sep software evolution
\end{keywords}

\maketitle

\section{Introduction}

In the contemporary digital landscape, software development plays a central role in driving innovation and functionality within technological systems. However, this process is not devoid of challenges, especially due to the increasing complexity of modern software systems. As software engineering evolves, the constant need to enhance developer efficiency and understanding becomes increasingly crucial. Critical information for understanding, such as control structures and dependencies, often gets lost in the intricate web of lines of code \cite{McConnell2004, martin2009}. This growing complexity underscores the importance of tools and approaches to alleviate developers' cognitive load. In response to this demand, the concept of ``Code Augmentation'' emerges as a promising approach to improve the interaction between developers and source code \cite{beelders2016syntax, SULIR2018}.

The ``Code Augmentation'' approach seeks to enrich the development experience by incorporating visual information directly into the source code. By adding visual elements such as colors, icons, and labels to source code, developers can gain faster and more accurate insights into the structure, flow, and relationships between pieces of code \cite{SULIR2018,beelders2016syntax}. Code augmentation, therefore, aims to create an environment where understanding the code is facilitated, allowing for a more intuitive and efficient interaction with software development.

The history of visual source code augmentation goes back to the early days of programming, when the first code editors introduced syntax highlighting features to improve readability and identify elements in code, called \textbf{identifiers} \cite{SULIR2018}.
The 1980s marked a milestone with the LEXX \cite{Cowlishaw1987} editor, which introduced real-time syntax highlighting. The evolution continues with advances, such as immediate feedback on syntax errors, as exemplified by the Magpie system \cite{Schwartz1984}. In the 1990s, projects such as ZStep95 \cite{Ungar1997} demonstrated interactivity by highlighting expressions under evaluation and displaying graphical output close to the source code. During the past two decades, with increasing computational power and advances in analysis algorithms, conventional IDEs have incorporated various augmenting features, encouraging research in this field \cite{SULIR2018, Murphy2019}.

Identifiers in the source code are ubiquitous: they make up around 70\% of a software system's source code size (measured in characters) and approximately a third of source code tokens \citep{deissenboeck2006concise}. Multiple studies have shown that descriptive variables, methods, or class names positively impact program comprehension: descriptive identifiers support program comprehension \citep{lawrie2006s,schankin2018descriptive}, shorter identifiers hamper comprehension \citep{hofmeister2017shorter} (although some specific ones might not \citep{beniamini2017meaningful}). Moreover, the identifier style and coding conventions can also have an impact on program comprehension \citep{binkley2009camelcase, sharafi2012women}. 

Refactorings \citep{fowler2018refactoring} are used to improve source code maintainability, where a large part of them are renamings \citep{hora2018assessing}. Many approaches have been proposed to improve identifiers, providing suggestions for renaming. 
These approaches range from normalizing source code identifiers \citep{guerrouj2013normalizing} to detecting naming bugs \citep{host2009debugging}, bad smells \citep{abebe2009lexicon}, or anti-patterns \citep{arnaoudova2016linguistic}. Other approaches leverage advanced algorithms or machine learning techniques to expand abbreviations \citep{lawrie2007extracting, corazza2012linsen}, provide renaming suggestions \citep{allamanis2015suggesting}, spot inconsistencies \citep{liu2019learning}, find bugs related to identifier names \citep{pradel2018deepbugs}, or even de-obfuscate source code \citep{vasilescu2017recovering}.

Beyond program comprehension, software identifiers, due to their predominance in software systems, are extensively used in Natural Language Processing (NLP) \citep{arnaoudova2015use}, and Machine Learning for Software Engineering \citep{allamanis2018survey} approaches to support a variety of software engineering tasks, including several of the renaming suggestions mentioned above (thus closing the loop), among many other approaches.

In the face of this scenario, this article aims to present a novel approach to code augmentation described as Impermanent Identifiers, aiming not only to enhance the software development experience but also to comprehensively analyze the adoption and acceptance of this technology in the development process.

The objectives of this study encompass two interrelated and equally significant aspects. First, we strive to make a substantial contribution to the field of code augmentation by proposing an approach centered around Impermanent Identifiers. This approach aims to explore the creation of dynamic identifiers that adapt to evolving contextual needs, enabling a more seamless and accurate interaction between developers and source code. In this regard, our research seeks to expand the concept of code augmentation, moving beyond conventional approaches and establishing the groundwork for an innovative methodology that enhances understanding, maintenance, and collaboration in the software development process.

Furthermore, our second objective is to conduct a rigorous evaluation of the adoption and acceptance of this technology in the software development landscape. Recognizing that the introduction of new technological approaches often encounters acceptance and integration challenges, a comprehensive empirical study was carried out. This involved assessing how developers perceive and adopt Impermanent Identifiers as part of their daily programming practices. We aimed to understand the perceived benefits, potential barriers, and factors influencing the decision to incorporate this innovative approach into their workflows.

In summary, this article seeks to set a new direction in the field of code augmentation by proposing the use of Impermanent Identifiers as a foundation, while simultaneously exploring the feasibility and acceptance of this approach within the developer community. Through this comprehensive and interdisciplinary research, we hope to contribute to the continuous improvement of software development practices and to the advancement of code augmentation as a whole.

Section \ref{background} provides the theoretical background used in this study, Section \ref{proposal} introduces the concept of Impermanent Identifiers, Section \ref{evaluation} outlines the methodology employed for the assessment of adoption and acceptance, Section \ref{results} presents the evaluation results, and finally, Section \ref{conclusion} concludes this work.  \section{Background}
\label{background}

This section provides fundamental background information on Code Identifiers (Section \ref{codeidentifiers}) and the UTAUT technique (Section \ref{UTAUT}), which was employed to assess technology acceptance.

\subsection{Code Identifiers}
\label{codeidentifiers}

Identifiers, often referred to as Code Identifiers, are fundamental elements within the realm of software development. They encompass variable names, method names, class names, and other symbols that developers use to label and reference various elements of their codebase. These identifiers are dynamic in nature, continually evolving as the software project progresses. They play a pivotal role in enhancing code readability, maintainability, and overall comprehension.

In software engineering, the choice of identifier names can significantly impact how effectively developers can understand and maintain a piece of code. Descriptive and meaningful identifiers improve program comprehension, making it easier for developers to grasp the purpose and functionality of different code components. Conversely, cryptic or overly brief identifiers can hinder comprehension and even introduce errors into the codebase \cite{arnaoudova2016linguistic}. Some programming languages, like C++, provide support for aliasing \cite{hackett2006aliasing}, when a data location in memory can be accessed through different identifiers in the program. While this allows different contexts to use different identifiers, this practice can also confuse developers \cite{arnaoudova2016linguistic}. 

Given their paramount importance, identifiers have been the subject of extensive research and tool development \cite{Petrescu2023}. Techniques and approaches have been devised to assist developers in choosing appropriate identifier names, managing their consistency, and even automating aspects of identifier maintenance. These efforts underscore the critical role that identifiers play in software development and the ongoing quest to optimize their usage for better code quality and efficiency.

In summary, we can assert that identifiers play two different roles: they function as a means of referencing a software entity and as a comprehensible description of its purpose for who is reading the code\cite{arnaoudova2016linguistic}. These roles often conflict because the identity must remain constant, as highlighted earlier, while the description presented to the programmer would benefit from being more adaptable, capable of varying over time and adjusting to diverse contexts. Thus, the first role of identifiers requires them to be permanently stored in the source code. However, for the second role identifiers could be impermanent and modified according to the context when being visualized by developers.

This paper advocates for the possibility of identifiers having textual representations different from their identity, which we call \textit{``impermanent identifiers''}. Freeing the textual representation from the constraints bound to the identity significantly reduces the cost of changing it. Based on this approach, the identifier can be adapted to different contexts, allowing the incorporation of additional information, omission of non-relevant parts, and its representation in another language. In addition, the temporary nature of impermanent identifiers allows them to leverage Unicode characters in a way that would not be practical for permanent identifiers. Using Unicode allows for concisely conveying context-sensitive information in the impermanent identifiers themselves.

For example, consider a Java method with the ``synchronized'' keyword, indicating that it can be accessed by only one thread at a time. While this information is relevant when developing or maintaining other classes that invoke this method, it is not explicit only by the method name. By using the concept of impermanent identifiers, the visualization of this method name could be changed (such as by adding an emoji) to make that fact explicit. Since the code is not changed, only the visualization, it does not affect any other code that depends on it.

Since the change affects only how the identifiers are visualized and not directly the code, it solves the issues above related to conflicts and changes in the code repository. The cost of renaming identifiers vanishing opens up new use cases for renaming identifiers with increasing frequency and also allows the surfacing of additional information in the identifiers. This practice also differs from aliasing since the identifier would be visualized consistently through the code, and two identifiers would not refer to the same variable simultaneously. Both can potentially improve program comprehension and evolution, allowing each developer to tailor how the code is visualized based on their needs.

Standard IDEs have altered the way they display source code in several ways. Different fonts and colors have been used for syntax highlighting for decades. Incorrect code is often underlined in red, while calls to deprecated methods can be displayed with a struck-out font. IDEs also display warnings or suggestions as clickable icons in the editor's left margin. However, the space available in the margin is limited, and the warning concerns the entire line.

IDEs have applied more changes to source code's appearance in recent years. This includes an inline ``rename refactoring'' interface that allows for quick (permanent) renaming of an identifier; nevertheless, this does not address the consequential implications of the refactoring process. Another example is source code folding, which selectively hides a section of source code (such as the body of an entire method or hiding exception handling code). Visual Studio Code\footnote{\url{https://code.visualstudio.com/} - 01/10/23} \citep{vscode} can display a ``Code Lens'', a textual line with additional information displayed in a smaller font above a method declaration. 

JetBrain's suite of IDEs \citep{jetbrains}  can show ``inlay hints''. Inlay hints are utilized to incorporate concise pieces of information immediately into the editor. This particular functionality offers enhanced code comprehension without inducing any interference with the workflow of the developer. A well-known example in this suite is the ``parameter hints'' in function or method calls: the IDE displays in smaller font the name of the parameter that will be bound to each call argument. Finally, Visual Studio Code's decoration API can display textual information on the right side of the code (such as who recently committed a line of code). 

Academic works have also tackled related problems. The Version Editor \citep{atkins2002using} used different fonts to show recently added and deleted code as far back as the 1980s. SubText \citep{edwards2005subtext} is a research tree-based programming language that also challenges the notion of textual identifiers. While Subtext is a new language, we aim for compatibility with existing languages. These examples illustrate how visual and block-based programming languages might not rely on textual identifiers. Instead, we explore these notions in more conventional text-based languages. \citet{Murphy-Hill2014} discusses factors that recommendations within IDEs must consider. Some academic IDE extensions surface information in the code editor. \citet{sliwerski2005hatari} displays the ``risk'' of a code section (obtained by querying software repositories) in the editor's margin. Code\_call\_lens extends Visual Studio's "Code Lens" with runtime and usage information \citep{janes2018code_call_lens}. 

The work by \citet{SULIR2018} presented a systematic mapping of source code editor augmentation tools and approaches. They reviewed 5553 articles published between 1998 and 2018 in two phases - keyword search and reference search. The result was a list of 103 relevant articles and a taxonomy of source code editor augmentation tools with seven dimensions:

\begin{enumerate}[topsep=0pt,itemsep=-1ex,partopsep=1ex,parsep=1ex]
    \item \textbf{Source:} Where does the data representing the augmentation come from?
\item \textbf{Type:} What data type does the augmentation represent?
\item \textbf{Visualization:} What does the augmentation look like?
\item \textbf{Location:} Where is the visual augmentation displayed?
\item \textbf{Target:} To what is the augmentation visually assigned?
\item \textbf{Interaction:} How can we interact with the augmentation?
\item \textbf{IDE:} For what IDE/editor is the augmentation implemented?
\end{enumerate}

\begin{figure}[!htb]
    \centering
    \includegraphics[width=15cm]{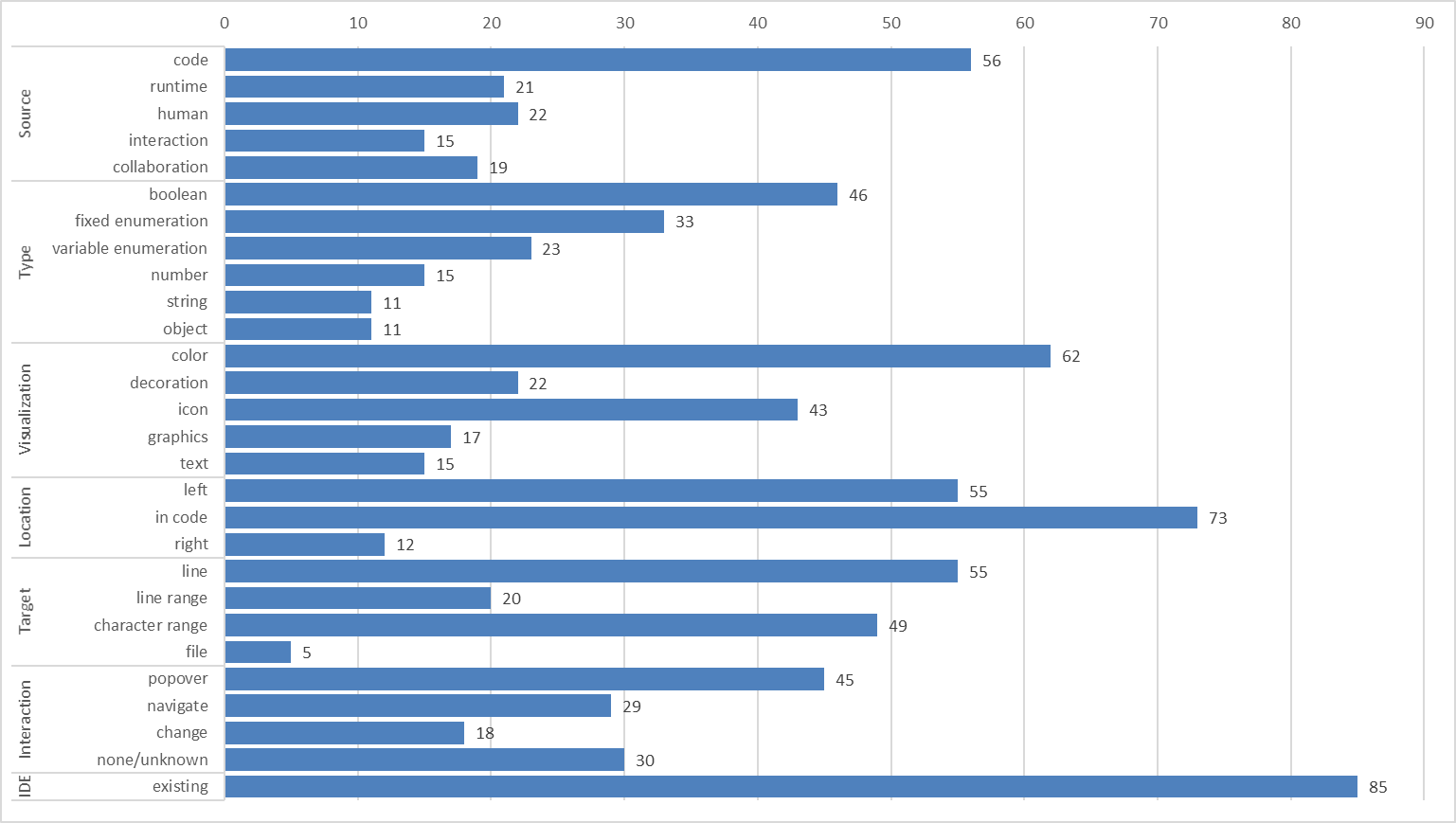}
    \caption{The numbers of tools fulfilling the given attributes of the dimensions \cite{SULIR2018}.} 
    \label{fig:RelatedWorks}
\end{figure}

In Figure~\ref{fig:RelatedWorks}, a chart illustrates how many reviewed tools satisfy each attribute within a specific dimension.

Most tools obtain their data through static source code analysis. The most common data category is ``boolean'', indicating the prevalence of binary information representation. The coloring technique is widely used for visualizing the enhancement. Most tools directly display the enhancement within the source code. The most frequent association for enhancement is with lines of code. Informational pop-ups are a common means of interaction with the enhancement. Most tools are implemented within existing IDEs. \subsection{Unified Theory of Acceptance and Use of Technology -- UTAUT}
\label{UTAUT}

This section describes the model we used to assess the adoption of code augmentation technology. 

The UTAUT (Unified Theory of Acceptance and Use of Technology) is an outcome of a comprehensive review conducted by \citet{Venkatesh2003} on various prior models that address the acceptance and utilization of technology. Models examined during the review include Theory of Reasoned Action (TRA), Technology Acceptance Model (TAM), Motivational Model (MM), Theory of Planned Behavior (TPB), Combined TPB and TAM (C-TPB-TAM), Model of PC Utilization (MPCU), Innovation Diffusion Theory (IDT), and Social Cognitive Theory (SCT). Through this meticulous review, the researchers integrated and unified the key components of these models to develop the UTAUT, which provides a comprehensive framework for understanding and predicting the behavior of technology adoption.

\begin{figure}[!htb]
    \centering
    \includegraphics[width=15cm]{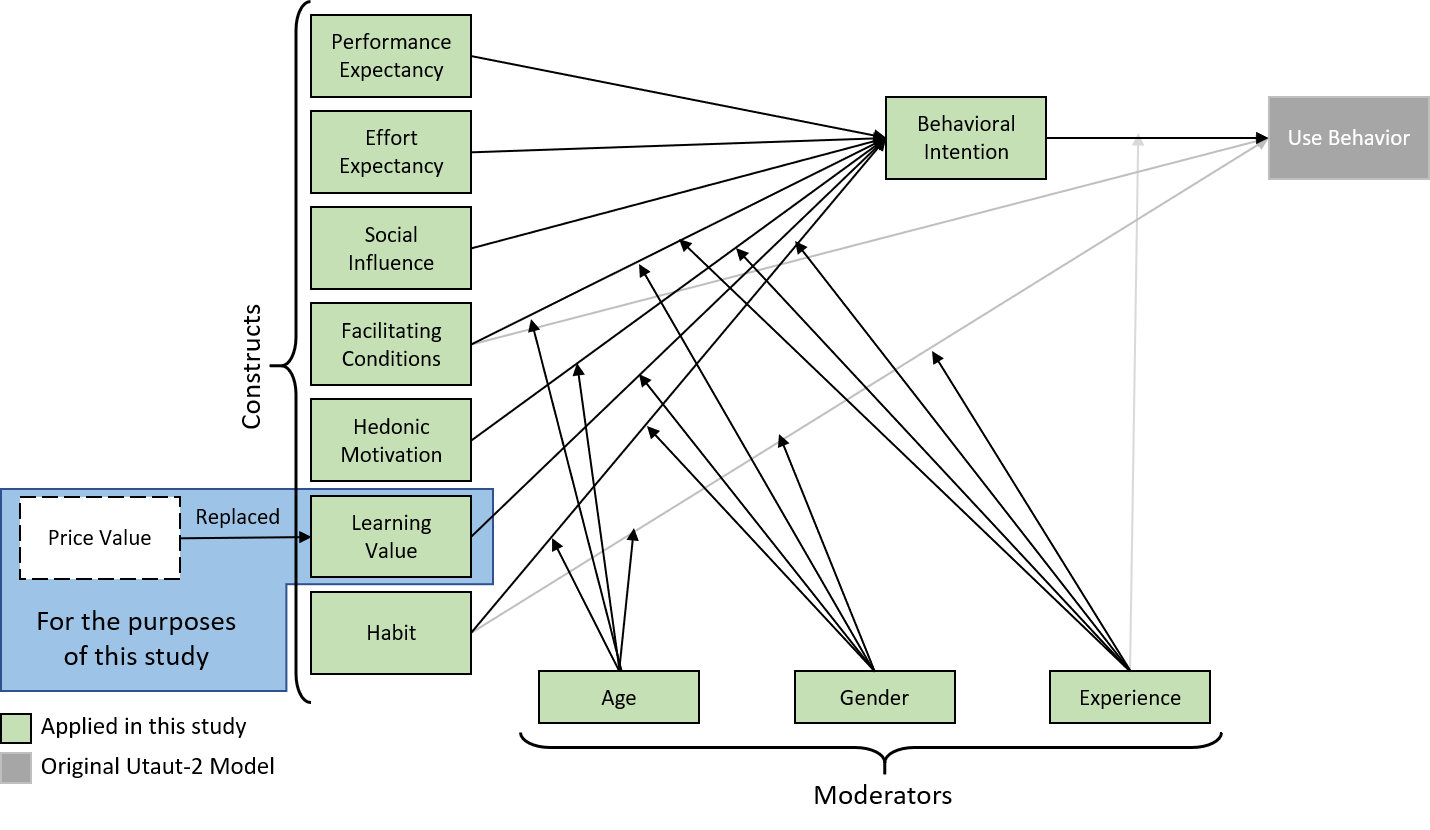}
    \caption{UTAUT 2 Model} 
    \label{fig:utaut}
\end{figure}

UTAUT 2 (Figure \ref{fig:utaut}) is a unified theory devised by \citet{Venkatesh2012}, serving as a comprehensive framework for examining individual technology adoption behavior. By extending and refining the original UTAUT model, UTAUT 2 further enhances the understanding of the factors influencing individual technology consumption.

As shown in Figure \ref{fig:utaut}, UTAUT 2 comprises seven distinct constructs, each playing a crucial role in understanding individual technology adoption:

\begin{enumerate}[label=\textbf{CT\arabic*}, leftmargin=25mm, topsep=0pt,itemsep=-1ex,partopsep=1ex,parsep=1ex]
    \item-- \textbf{Performance Expectation:} This construct relates to how individuals perceive the extent to which a particular technology enhances task performance.
    \label{CT-PerfExp}
    \item-- \textbf{Expectation of Effort:} Referring to the level of ease or difficulty experienced in using the technology.
    \label{CT-ExpEff}
    \item-- \textbf{Social Influence:} This construct assesses the significance individuals attach to the technology's use based on its importance to others.
    \label{CT-SocialInf}
    \item-- \textbf{Facilitating Conditions:} Characterized by the availability of support and resources for technology use.
    \label{CT-FaciCond}
    \item-- \textbf{Hedonic Motivations:} Linked to the pleasure, enjoyment, and interest individuals associate with adopting technological innovations.
    \label{CT-Hedonic}
    \item-- \textbf{Learning value:} This construct involves the individual's Learning Value, which refers to the value they place on education and their perception of its cost-effectiveness.
    \label{CT-LeanVal}
    \item-- \textbf{Habit}: Ascribing meaning to routine, custom, and behavioral patterns.
    \label{CT-Habit}
\end{enumerate}

These constructs serve as independent variables in the UTAUT 2 model, with the dependent variables being Behavioral Intention and, subsequently, Use Behavior. Moreover, gender, age, and experience play a moderating role in this theoretical framework.

The moderators in the UTAUT 2 are additional factors that influence the relationship between the independent variables (constructs) and the dependent variables (Behavioral Intention and Use Behavior). These moderators help to explain how individual differences can impact the technology adoption process. In the UTAUT 2 model, the following moderators are considered:

\begin{enumerate}[label=\textbf{MT\arabic*}, leftmargin=25mm,topsep=0pt,itemsep=-1ex,partopsep=1ex,parsep=1ex]
    \item-- \textbf{Gender:} The effect of gender differences on technology adoption is examined to understand whether certain technologies are more likely to be adopted by one gender over another.
    \item-- \textbf{Age:} Age-related differences in acceptance and use of technology are analyzed to explore how age influences individuals' willingness to adopt and use technology.
    \item-- \textbf{Experience:} The level of prior experience and familiarity with similar technologies is investigated to determine its influence on individuals' acceptance and use of the new technology.
\end{enumerate}

These moderators provide insights into how various demographic and experiential factors may interact with the core constructs to influence an individual's behavioral intention and actual use of technology.

\section{From Code Augmentation to Impermanent Identifiers}
\label{proposal}

The evolution of software development tools, such as IDEs (Integrated Development Environments), has brought numerous improvements aimed at making the programming process more efficient and collaborative, especially in the case of code augmentation. However, as IDEs have become increasingly powerful, a common characteristic has persisted: the lack of comprehensive customization for code augmentation functionalities. Many IDEs implement features in a general way, providing little flexibility for developers to tailor the tools to their specific needs. This becomes particularly evident in critical areas, such as code manipulation and team collaboration.

A clear example of this phenomenon is the refactoring of identifiers, such as the renaming operation. Although it is a crucial functionality for improving code readability and maintenance, most IDEs implement this operation in a way that directly affects all team members. Renaming an identifier can have significant implications, and these changes are often uniformly reflected throughout the project, without considering individual preferences or different interpretations of the same code.

This situation raises a fundamental question: is there an alternative approach that allows developers to have more refined customization of their programming experiences while preserving integrity and team collaboration? The answer to this question leads us to explore impermanent identifiers. This approach decouples the textual representation of an identifier from its identity, enabling deeper customization without the complications associated with physical changes to the source code.

\subsection{Challenge: Identifiers are hard to change}

While an identifier can convey a lot of very useful information to developers, this is not its technical function. The technical function of a software identifier is to \emph{identify} the specific software element it refers to, be it a variable, a function, a method, a class, a package, a module, etc. A computer that executes a program can completely disregard the identifiers themselves, which is why computer-generated code often uses meaningless identifiers.

In text-based languages (the most widely used in practice), the two properties of identifying a software element and describing its purpose to developers are conflated. It makes renaming an identifier more difficult than it should be, as it involves changing all textual occurrences of the identifier to the new name. Moreover, different identifier approaches might be more suitable for developers with diverse backgrounds and roles. For instance, shorter identifiers might make reading more straightforward for those who already know that part of the code, but more descriptive ones might improve the understanding of first-timers. 

Even if several IDEs automate the rename refactoring, the execution of manual steps and the presence of naming conflicts can lead to incomplete refactorings and bugs. In some cases, renaming an identifier to an alternative that is more suitable for one developer might make it worse for another\footnote{A famous quote says \citep{fowler2009}: ``There are only two hard things in Computer Science: cache invalidation and naming things''}, as in multilingual codebases. In other cases, the refactoring is not accessible to the developer or might affect external code. If the identifier comes from an external library, it cannot be changed by the code that uses it. In this case, an alternative can be to create an adapter that provides access to external services under new names. From the perspective of the API designer, changing a name can potentially break the client's code. Even with mechanisms such as deprecation,  evidence shows that they are not very successful, as clients often do not migrate or do it after a very long time \citep{robbes2012developers, sawant2016reaction}.

Even successful rename refactorings might cause inconvenience, such as causing disorientation in developers used to the previous alternative. This impact can also be noticed in the versioning system, especially in large-scale text-based renaming: it generates large commits that might tangle and conflict with other changes. This affects tools and empirical studies that evaluate commits since they might overestimate the importance of a commit or lose track of a source code entity \citep{hora2018assessing}. 

All of these factors make renaming an identifier a decision that should not be taken lightly. Thus, developers may avoid renaming opportunities that could benefit code comprehension due to their cost.
\citet{Arnaoudova2014} surveyed 71 developers on their renaming habits. They found that only 8\% of developers said that renaming is straightforward. Regarding cost, most surveyed developers answered that even with the help of automated tools, renaming can have a significant cost (particularly for large-scale renames or ones that break client compatibility). \subsection{Proposal: Impermanent identifiers}

This work aims to investigate the feasibility of adapting and adjusting impermanent identifiers according to the specific needs of different software projects or programming contexts. This involves creating a flexible approach that allows developers to customize and shape impermanent identifiers effectively, aiming to enhance code comprehension, facilitate the development process, and promote more efficient and sustainable software maintenance.

In the following subsections, we provide a non-exhaustive list of scenarios in which ``Impermanent identifiers'' could be useful.

\subsubsection{Impermanent identifiers facilitating aliasing}

One particularly interesting aspect of impermanent identifiers is that they enable \textit{aliasing}, i.e., to choose a \emph{personal name} for an identifier. In other words, a developer might choose to ``rename'' identifiers for their own use while not affecting the identifier that everyone else sees. Additionally, the metadata from these personal renaming operations might be shared with other team members, creating name profiles that allow switching between different ways of visualizing the code.

Another important application of aliasing is that it can also be applied to identifiers that come from external code, such as libraries and APIs. This allows the adaptation of the terminology used in external classes and methods to the local context of a particular application. It also allows an API designer to provide alternative names for different user profiles or use cases.
For instance, a user that uses the default settings of an API can use a name profile with simple and direct names, while a more advanced user might see more descriptive names that reveal details of the internal behavior.

\textbf{Automated suggestions.} These alias suggestions could come either from automated algorithms or be manually suggested by developers. An instance of automated aliasing can be the change between alternatives for naming conventions, like \texttt{camelCase}, \texttt{snake\_case}, or others, such as Hungarian Notation. Automation could also abbreviate and expand names according to certain rules. On the one hand, the abbreviation can be used to reduce the amount of reading and visual overload from the code. One example could be to simplify the name of a \texttt{VeryLongJavaLanguageException} to \texttt{VLJLException} or even just \texttt{VLJL}{\Large\texttwemoji{stop_sign}}{\Large\texttwemoji{shorts}}, with a ``stop'' Unicode character denoting an exception, and decorated with a ``shorts'' Unicode character indicating that the name has been shortened\footnote{Please note that the precise choice of emojis is not the point of the paper and could even be personalized by each developer or even by domain applications.}. On the other hand, expansion can help developers unfamiliar with a certain area of the code in its comprehension. Adding into the name information about parameters and returns can make its functionality more explicit. For instance, a method named \texttt{getUsers()} could be automatically expanded to  \texttt{getUsersByStatus()} by adding information about the arguments. Some code annotations can also carry some information that might be useful to be present in the identifier name \citep{guerra2016annotations}, such as the ones that can be used to identify the class role or configure additional method behavior (e.g., transactions, access control).

\textbf{Manual suggestions.} Manual aliasing could also be helpful in all the cases mentioned in the previous paragraph. The trade-off is between the possibility of automatically aliasing several identifiers quickly and the capability to add semantic information that could not be inferred through the code. A scenario that could combine both approaches is the translation of the identifiers from an API to a different natural language spoken by some developers. While an initial translation of terms present in the identifiers can be done by an automated tool, manual adjustments can be used to tailor some names to more suitable alternatives related to the application domain.

\textbf{Supporting change processes.} Supporting systematic change processes, such as impact analysis and change propagation, is another application of manually inserted impermanent identifiers. These processes ensure that changes are thoroughly performed by maintaining a list of software entities that need to be checked for changes, marking them with status (e.g., inspection pending, needs to change, no change needed, changes implemented). This list is usually maintained either in a separate tool (e.g., a spreadsheet) or in a dedicated IDE plugin (e.g., JRipples \citep{buckner2005jripples}). Impermanent identifiers might straightforwardly support such a process by simply aliasing the identifiers concerned to show their current status after their name. Since modern IDEs support Unicode, this status can be encoded in a single character (e.g., inspection pending {\Large\texttwemoji{orange_circle}}, needs to change {\Large\texttwemoji{red_circle}}, no change needed {\Large\texttwemoji{green_circle}}), changes implemented {\Large\texttwemoji{purple_circle}}). Accordingly, a status change is immediately followed by an action to generate the updated alias for the identifier. 

\textbf{Aliasing vs renaming.} Technically, all these applications are possible with the classical rename refactoring (except when applied to library code and modulo each programming language's support for Unicode). However, in scenarios where the renaming is contextual or personal, they would likely need to be undone before committing code to avoid interference. In practice, this considerable friction makes applying code refactoring for these scenarios unfeasible. Moreover, the possibility to quickly switch between different impermanent identifiers can be used to adapt the code visualization for the task being performed. 

\subsubsection{Impermanent identifiers surfacing information}
\label{ImpIden}

Beyond facilitating aliasing, impermanent identifiers make surfacing information more immediate and straightforward by temporarily appending relevant information to the identifier name. The additional information can be textual or, when leveraging Unicode, be significantly more compact.

The information could come from any analysis tool (e.g., source code, software repository). It can be used not only to surface important information while a programmer looks at the identifier definition but also when a programmer encounters a usage of the identifier in the code base. This information may help the user decide if they need further inspecting a software entity's declaration or implementation. Similar mechanisms exist in IDEs, particularly by showing information (warnings, quick fixes, etc.) in the margin of the code. However, while these are limited to information that applies to the entire line, impermanent identifiers can show multiple pieces of information associated with single or multiple identifiers.

\textbf{In conjunction with aliasing.} In the context of aliasing, an analysis could evaluate if an identifier name has quality issues and surface the result as a part of it (e.g., {\Large\texttwemoji{see_no_evil}}, {\Large\texttwemoji{nauseated_face}}, {\Large\texttwemoji{vomiting_face}}, or {\Large\texttwemoji{poop}}, depending on the naming issue's severity). The developers would then be prompted to investigate alternative names manually or via a suggestion algorithm. After a recent rename, the refactoring can be shared with the entire development team, and a symbol indicating this status could be appended to the identifier, such as {\Large\texttwemoji{writing hand}} or {\Large\texttwemoji{new}}. That could mitigate the disorientation associated with renaming by pointing out which identifiers were renamed or aliased, possibly showing the previous name. Since identifiers can be personalized, a developer might even elect to keep the old name as their version.

\textbf{Surfacing additional properties.} However, the approach is more general and can be applied beyond aliasing. An example is presenting static properties of source code entities that are visible only when declared but not when used. This could lead to less source code navigation and fewer context switches, affecting productivity \citep{meyer2014software, sanchez2015empirical}. Examples include visibility modifiers (e.g., \texttt{private}{\Large\texttwemoji{locked}}, \texttt{public}{\Large\texttwemoji{unlocked}}), asynchronous (e.g., \texttt{async}{\Large\texttwemoji{hourglass not done}}), and nullability (e.g., null {\Large\texttwemoji{red question mark}}). In some JavaScript libraries, asynchronous method or function calls end conventionally with the suffix ``Async'', such as \texttt{Camera.takePictureAsync()}. An indicator of the presence of some code annotations might also be useful since they can reveal a class architectural role (e.g., persistent entities {\Large\texttwemoji{floppy disk}}, controllers {\Large\texttwemoji{shuffle tracks button}}) or method modifiers (e.g., access control {\Large\texttwemoji{police_officer}}, callback conditions {\Large\texttwemoji{right arrow curving left}})\citep{lima2022anotarch}.

The surfaced information can also come from external sources, like analysis tools or the code repository. For instance, according to an analysis of the source code repository, source code entities that have recently changed could display this information (e.g., {\Large\texttwemoji{construction}}). This could help focus code review activities, and small teams might even have an identification of \emph{who} changed it if each team member picks a specific character. Other repository analyses could add information about calls to code that are known to be risky \citep{sliwerski2005hatari} ({\Large\texttwemoji{warning}}) or that show signs of decay ({\Large\texttwemoji{derelict house}}). The information can be shown in the identifier declaration and its usage for that application.

Needless to say, static or dynamic analyses (e.g., FindBugs \citep{hovemeyer2004finding}, etc.) could also be integrated in this fashion. There are a variety of possible indicators of different warnings with several levels of severity (e.g., {\Large\texttwemoji{fire}}, {\Large\texttwemoji{police car light}}, {\Large\texttwemoji{skull and crossbones}}). Advanced static analysis tools such as Infer \citep{calcagno2011infer} could be leveraged to surface information such as a method's estimated complexity or side effects. Also, an analysis of the energy consumption of components \citep{Corral15} or APIs \citep{hasan2016energy} could be used to annotate software elements with high energy consumption (e.g., {\Large\texttwemoji{battery}}, {\Large\texttwemoji{electric plug}}).

\textbf{System-specific adaptations.} Additionally, system-specific rules can also be implemented at the team and individual levels to better reflect project-specific coding conventions. API providers could use impermanent identifiers to highlight usage patterns of their libraries (e.g., to highlight that a call to \texttt{open} {\Large\texttwemoji{open book}} should be followed by a call to \texttt{close} {\Large\texttwemoji{closed book}}). 

\textbf{Dealing with information overload.} Clearly, excessive proliferation of such annotations may have adverse effects. Each identifier annotation should be easily enabled/disabled to customize the environment by each team or user according to each context or activity. For instance, users could selectively enable annotations that are related to code quality issues when they are reviewing code. They may enable annotations related to performance when they are optimizing code, or enable annotations displaying properties such as visibility, architectural roles, when doing code comprehension tasks. Focusing on a specific category of annotations instead of all of them should already reduce the amount of annotations visible at any time, as well as making the annotations more semantically homogeneous. We also envision ordering the surfaced annotations by user or team-defined priority and providing a ``slider'' to quickly regulate the amount of information that is visible at a glance (e.g., displaying only high-priority annotations by default and adding additional information on demand).

\subsubsection{Implementation examples}

\begin{figure}[ht]
    \centering
    \includegraphics[width=.8\textwidth]{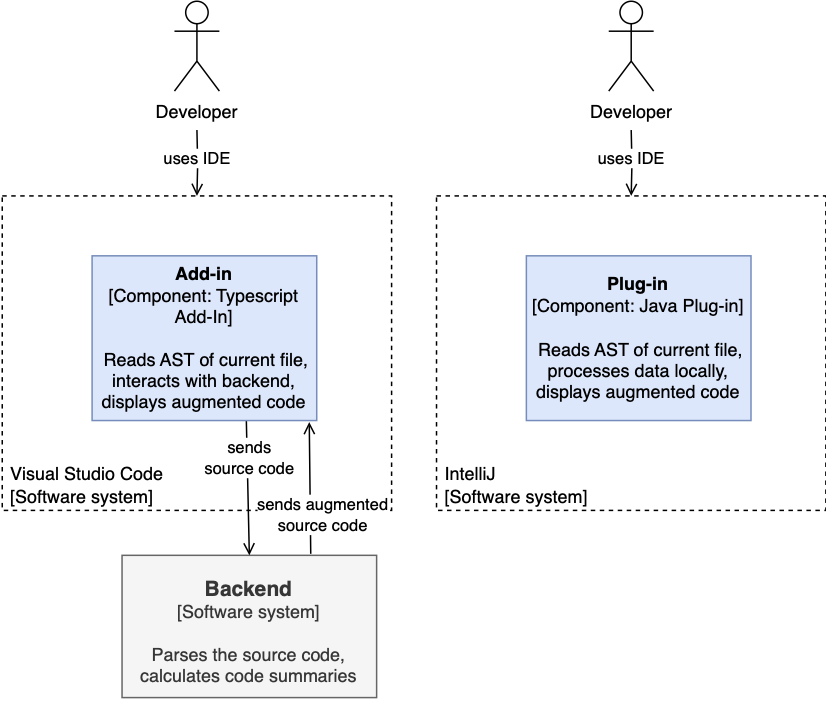}
    \caption{Component diagram (following the C4 notation) \cite{Brown2023} of the first (on the left) and second (on the right) proof-of-concept.} 
    \label{fig:poc1}
\end{figure}

We created two proof-of-concept implementations in different IDEs to show the feasibility of implementing impermanent identifiers. The first implementation is an IDE extension for Visual Studio Code called "Adverb"\footnote{\url{https://github.com/lalenzos/ar}} was written in TypeScript and was used to experiment with different options to augment code, e.g., the display of variables in snake\_case (while being written in camelCase) or vice versa; also the display of inline parameters was tested (i.e., the visualization of a function {\tt{add(2, 5)}} as {\tt{add(\textcolor{darkgray}{argument1: }2, \textcolor{darkgray}{argument2: }5)}} in the editor, obviously without changing the underlying source code. The architecture of this first proof of concept is depicted in Fig. \ref{fig:poc1} on the left: since code summaries required the use of a large language model, these operations took place on the backend, while the add-in\footnote{In Visual Studio Code, extensions are called ``add-ins'', in IntelliJ, they are called ``plug-ins''. We follow this convention in our description.} was responsible for the visualization of the results.

In the future, we do not forsee that there will be a dedicated backend for the plug-in, but there will be interaction with external systems like vulnerability databases, source code quality tools like SonarQube, etc.

The described architecture made it difficult to progress with the project: two parts had to be kept in synch and the use of large language models required to be very attentive not to slow down the developer experience. The source code of the first proof-of-concept is still available at the above mentioned link but was not pursued any further. The second proof-of-concept implementation was developed with a more lightweight architecture in mind, i.e., as a plugin for JetBrains IntelliJ Idea, named ``codEEmoji'', which provides support for the Java programming language. Its architecture is depicted in Fig. \ref{fig:poc1} on the right: a single plug-in reads the abstract syntax tree of the current source code file, processes the data and displays the augmented code. The codEEmoji plugin is also available on the JetBrains IntelliJ Marketplace\footnote{\url{https://plugins.jetbrains.com/plugin/22416-codeemoji/}} and can be installed directly through the IDE.

The Visual Studio Code extension "Adverb" locates the identifiers to be renamed by traversing the current file's Abstract Syntax Tree. The extension maintains a list of mappings from impermanent identifiers to their permanent counterparts in a project-level file. If the extension encounters an identifier that should be aliased to an impermanent identifier, it uses Visual Studio Code's Decoration API \cite{vscodeapi} to visually replace the permanent identifier with the impermanent one. 

The extension provides several semi-automated aliasing strategies (such as abbreviating identifiers, changing coding conventions, and using Neural Language Models such as CodeBERT \cite{feng2020codebert} for renaming suggestions), as well as allowing fully manual renames. Initial tests show that the proof of concept scales appropriately, even for larger files with a large number of impermanent identifiers. Currently, the proof-of-concept limitations are that impermanent identifiers are disabled when a line is being edited (i.e., the line where the cursor is located) and that we support only Javascript and TypeScript at the moment.

\begin{figure}[h!]
    \centering
    \includegraphics[width=.58\textwidth]{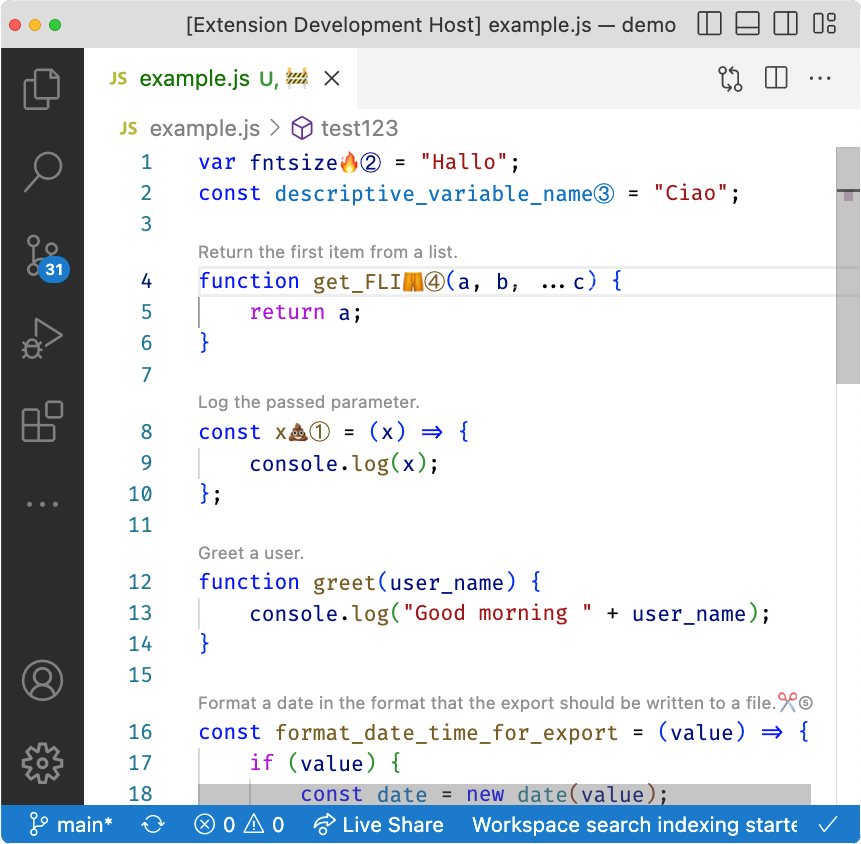}
    \caption{A screen capture of our VSCode extension.} 
    \label{fig:demo1}
\end{figure}

A screen capture of our VSCode extension "Adverb" is shown in Figure~\ref{fig:demo1} with several impermanent identifiers: an identifier with \ding{192} a naming issue; \ding{193} a source code warning; \ding{194} an expanded description; \ding{195} an algorithm-generated abbreviation; \ding{196} a code lens description that suggests to shorten a method. Furthermore, in this scenario, the developer opted to display identifiers using a \texttt{snake\_case} style, and we show that the impermanent identifiers are active decorating the file name with {\Large\texttwemoji{construction}}.

\begin{figure}[h!]
    \centering
    \includegraphics[width=.58\textwidth]{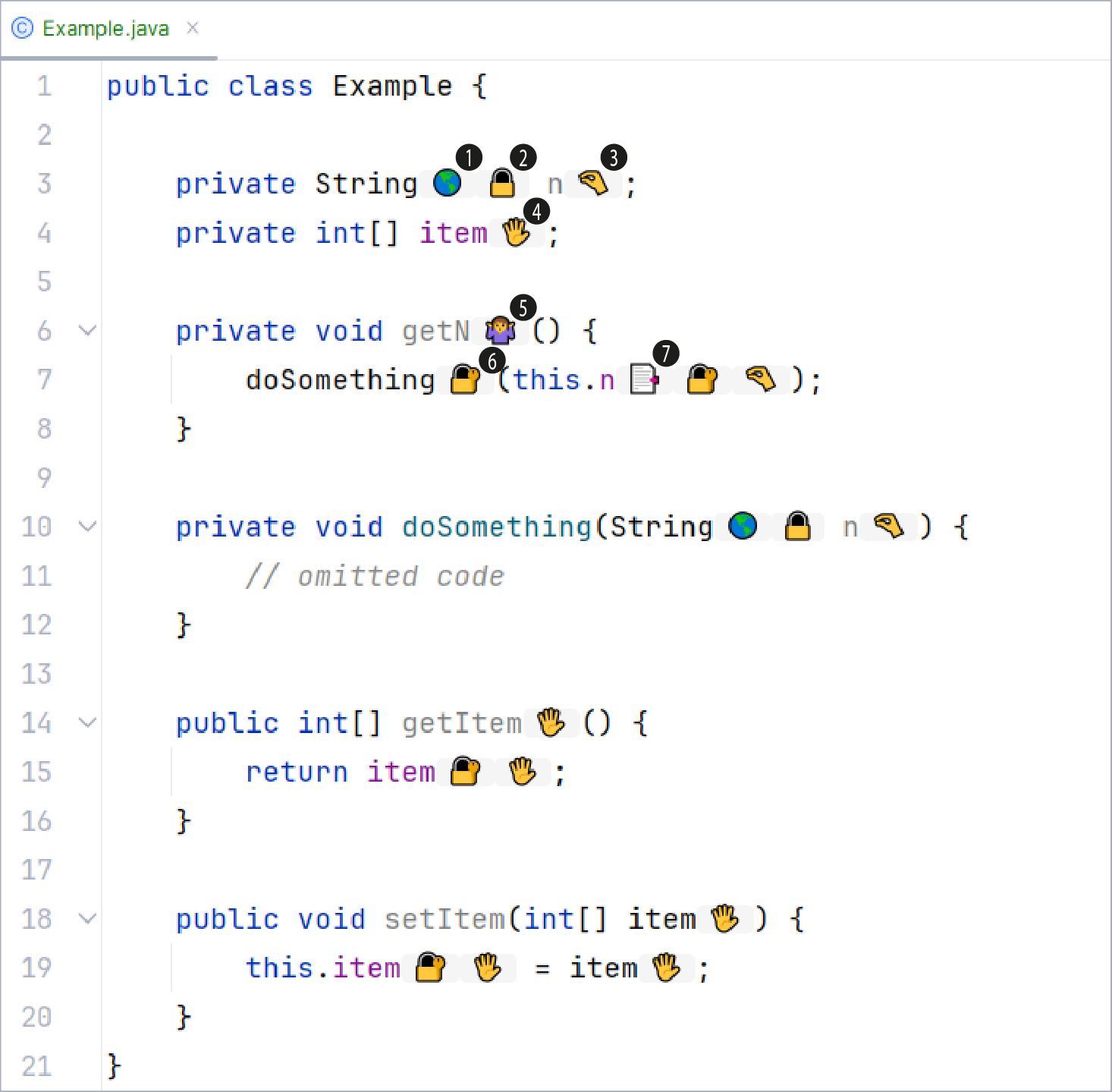}
    \caption{A screen capture of our IntelliJ Idea plugin (codEEmoji).} 
    \label{fig:demo2}
\end{figure}

In the codEEmoji plugin, the emojis are used to represent cases involving anti-patterns, such as violations of naming conventions or bad programming practices \cite{arnaoudova2016linguistic}. The plugin prototype also displays emojis to represent the use of modifiers in classes, fields, methods, parameters, or local variables. Similarly, the developer has the ability to specify the emoji that will be shown for each of these items based on a predefined rule. For example, emojis can be assigned to a class that is being instantiated and implements a specific interface, a method that is invoked and annotated with a specific annotation, and a local variable of a specific type. 

The codEEmoji plugin is under to the terms and conditions of the GNU General Public Licence version 3.0. Using the API for Inlay Hints from the Intellij IDE\footnote{\url{https://plugins.jetbrains.com/docs/intellij/inlay-hints.html}}, the plugin was developed such that it could function as an extensible framework. It is possible for developers to gain more code understanding with the use of inlay hints, which render small amounts of information right into the editor without disruption to the workflow. Emojis are automatically added to the code during coding time when the code is being developed. All details about the implementation  can be viewed at the address \url{https://github.com/codeemoji/codeemoji-plugin}.

A screen capture of the codEEmoji plugin is shown in Figure~\ref{fig:demo2} with several impermanent identifiers: an identifier with \ding{192} a class has public modifier; \ding{193} class has final modifier; \ding{194} a naming issue (short descriptive name); \ding{195} a naming issue (says one but contains many); \ding{196} bad conventions for method (getter does not return); \ding{197} method has private modifier; \ding{198} specific emoji for project (fields of String type).

With IDE support, implementing impermanent identifiers does not affect source code parsers. Our current IntelliJ Idea and Visual Studio Code implementations do not change the (permanent) identifiers in the source code. The identifiers are only modified when the code is displayed to the user in the visualization and not directly in the source code; all the IDE tools, such as linters, static analyzers, and code completion, that work on the original code do not need any modification whatsoever. It is important to highlight that this property depends on the specific IDE: for Visual Studio Code, we use the Code's Decoration API; for IntelliJ idea, we use the Inlay Hints API. If an IDE lacks an API that allows us to implement the extension unobtrusively, an approach that modifies the code would significantly hamper the approach's applicability.

The two presented plugins are currently in the experimental stage. Nevertheless, both serve as evidence of the practical viability of the impermanent identifiers. In both cases, it was possible to show the code with the impermanent identifiers without actually changing the code. 

  \section{Adoption and Acceptance Evaluation}
\label{evaluation}

\subsection{Objective}

After introducing the concept of ``code augmentation'', substantial work remains to be done. The next phase involved evaluating the acceptance and adoption of this new technology. The evaluation focused on understanding how the described scenarios impacted program comprehension and software evolution tasks. To ensure a comprehensive analysis, individual scenario assessments were conducted first, followed by combined scenario assessments to examine the potential effect of information overload.

In this section, we present the evaluation of the adoption and acceptance of the code augmentation technology. We developed a survey based on the UTAUT 2 methodology to conduct this evaluation.

In the context of this study, ``Price Value'' was replaced by ``Learning Value'' since technology consumption does not necessarily involve monetary acquisition. However, code augmentation can contribute to the learning process, so ``Learning Value'' was included as a relevant construct.

\subsection{Survey Design}
\label{surveydesignsection}
A comprehensive survey in the form of a questionnaire was meticulously designed and encompasses both open-ended and closed-ended questions to evaluate the degree of adoption and acceptance of ``code augmentation''.

The survey consisted of five distinct modules. Commencing with the initial module, participants were required to confirm their consent to participate and agree to the survey's terms. Subsequently, in the second module, participants were prompted to provide their data, sociodemographic information, and technical profiles.

Moving forward to the third module, participants were presented with crucial survey instructions and asked to indicate their clarity regarding them. The fourth module introduced 11 application and usage scenarios to assess user acceptance using the UTAUT 2 concept and framework. Finally, the fifth module offered participants the opportunity to express their preferences and perceptions through open-ended questions that sought to elicit overall impressions regarding the code augmentation presented. In this particular case, the participant had the freedom to express their preferred form of feedback, including recommendations, compliments, evaluations, and observations.

The fourth module was purposefully crafted to evaluate 11 distinct code augmentation usage scenarios. These scenarios were designed to gauge varying acceptance levels within each context, recognizing that the benefits of code augmentation technology can differ based on its application context. By assessing the technology's reception across diverse scenarios, we aim to gain insights into its potential impact and effectiveness in different settings.  The scenarios considered are described (code examples of each one are available in the survey text, as part of the replication package):

\begin{enumerate}[label=\textbf{SC\protect\twodigits{\theenumi}}, leftmargin=25mm,topsep=0pt,itemsep=-1ex,partopsep=1ex,parsep=1ex]
    \item-- \textbf{Showing more \underline{descriptive names}:} Code augmentation can be used to display more descriptive names for methods and classes in their usage, based on the information extracted from their declarations. For example, the methods ``addTranslation()'' and ``translate()'' can be expanded to show additional information based on their parameter names.. 
    \label{sc1}
    \item-- \textbf{Showing \underline{abbreviated names}:} This code augmentation technique can employ rules to shorten method and class names as they appear in the code. In this example, the prefix ``get'' can be removed from methods, and the names of exception classes can be simplified in the ``catch'' clause, transforming, for instance, ``InvocationTargetException' into ``InvTarExp''. Furthermore, the names of well-known method calls in Java, such as ``System.out.println()'' and ``printStackTrace()'', can also be simplified.
    \label{sc2}
    \item-- \textbf{Variables with \underline{bad naming} practices:} This code augmentation showcases variables using improper naming practices that can be enhanced with emojis representing their issue. The emoji {\Large\texttwemoji{input numbers}} can signify a variable for storing multiple elements, even though it has a singular name. The emoji {\Large\texttwemoji{pinching hand}} can label a variable with just a single letter, lacking meaningful naming.
    \label{sc3}
    \item-- \textbf{\underline{Recent renames}:} This case uses code augmentation to add the emoji {\Large\texttwemoji{writing hand}} to a variable name to represent that this variable was recently renamed. 
    \label{sc4}
    \item-- \textbf{\underline{New} methods and \underline{updated} methods:} The code augmentation add the emoji {\Large\texttwemoji{baby}} to methods that were recently added in the code and the emoji {\Large\texttwemoji{pencil}} in methods that were recently changed. In this scenario, the emojis are added to the method declaration.
    \label{sc5}
    \item-- \textbf{\underline{Showing} the presence of \underline{modifiers} in method usage:} The code augmentation adds the emoji {\Large\texttwemoji{vertical traffic light}} to a method name to indicate that it has the ``synchronized'' modifier. 
    \label{sc6}
    \item-- \textbf{\underline{Showing} the presence of \underline{annotation} on method and class usage:} The code is augmented to represent the transaction attribute of a method when it is invoked. The emojis {\Large\texttwemoji{new}}, {\Large\texttwemoji{right arrow}}, and {\Large\texttwemoji{no entry}} represent respectively the configurations ``REQUIRES\_NEW'', ``REQUIRES'' and ``NOT\_SUPPORTED'' for the @TransactionAttribute annotation.
    \label{sc7}
    \item-- \textbf{Calls to \underline{risky code}:} This code augmentation to add the emoji {\Large\texttwemoji{skull and crossbones}} to a method call from a library that can bring known security risks to the code (identified by an external tool).
    \label{sc8}
    \item-- \textbf{\underline{Problems detected} through static analysis tools:} The code augmentation can highlight a bad practice where an exception is printed and not re-thrown, potentially leading to the continuation of execution in a situation where it should be halted. In such cases, an emoji {\Large\texttwemoji{woman shrugging}} is added to the variable name representing the `lost' exception.
    \label{sc9}
    \item-- \textbf{\underline{Visual reminders} on API usage:} Code augmentation can be used to add reminders, such as providing a visual cue when a PrintWriter instance needs to be closed after usage. In this case, the variable that holds the created instance is annotated with the emoji {\Large\texttwemoji{open book}} to indicate that it is in an open state, and the method used to close it is marked with the emoji {\Large\texttwemoji{blue book}}.
    \label{sc10}
    \item-- \textbf{Showing the last \underline{person} that \underline{updated} some code:} In this scenario, code augmentation can be employed to identify the most recent developer who modified a specific method. Here, the {\Large\texttwemoji{girl: medium-dark skin tone}} and {\Large\texttwemoji{cowboy hat face}} emojis are used to symbolize developer avatars.
    \label{sc11}
\end{enumerate}

In this module (fourth) the survey was designed according to the UTAUT 2 methodology, wherein closed-ended questions were specifically formulated. The closed-ended questions are formulated as declarative statements and employ Likert's unidimensional scale for the response options, prompting the participant to indicate their level of agreement on a continuum ranging from ``Strongly Agree'' (denoted as 5) to ``Strongly disagree'' (denoted as 1). For each scenario, 16 questions were applied (Table~\ref{table:ClosedQuestions}), which were classified according to the constructs predicted by UTAUT 2 (Figure~\ref{fig:utaut} and Figure~\ref{fig:UtautHip}). 

\begin{table}[width=.9\linewidth,pos=!htb]
    \begin{center}
    \caption{Closed-ended questions were formulated as declarative statements.}
    \label{table:ClosedQuestions}
\resizebox{16.4cm}{!}{
        \begin{tabular}{c p{9.7cm} c c}
            \toprule
                \multicolumn{1}{c}{\textbf{\#}} & \multicolumn{1}{l}{\textbf{Declarative statements}} & \multicolumn{1}{c}{\textbf{UTAUT 2}}  & \multicolumn{1}{c}{\textbf{Technical application}}\\
            \midrule
                01 & This code augmentation would improve my performance in creating new code. & \ref{CT-PerfExp} & New Code \\
                02 & This code augmentation would improve my performance in understanding and reviewing code. & \ref{CT-PerfExp} & Code Review \\
                03 & This code augmentation would improve my performance in refactoring and modifying code. & \ref{CT-PerfExp} & Code Refactoring\\
                04 & This code augmentation would make it easy to create new code. & \ref{CT-ExpEff} & New Code\\
                05 & This code augmentation would make it easy to understand and review code. & \ref{CT-ExpEff} & Code Review\\
                06 & This code augmentation would make it easy to refactor and modify code. & \ref{CT-ExpEff} & Code refactoring\\
                07 & This code augmentation would make my team work better together. & \ref{CT-SocialInf} & Team Efficiency\\
                08 & If other team members use this code augmentation, that would make me want to use it too. & \ref{CT-SocialInf} & Team Efficiency\\
                09 & It would facilitate the application of this code augmentation if it is available in my IDE. & \ref{CT-FaciCond} \\
                10 & It would facilitate the application of this code augmentation if it is available in web tools to browse the source code, like code repositories and continuous integration & \ref{CT-FaciCond} \\
                11 & Using this code augmentation would make me learn new things. & \ref{CT-LeanVal}\\
                12 & Using this code augmentation would make me consolidate the learning of things I already know. & \ref{CT-LeanVal}\\
                13 & It would be pleasant to work with the code using this code augmentation. & \ref{CT-Hedonic}\\
                14 & It would be more fun to work with the code using this code augmentation. & \ref{CT-Hedonic}\\
                15 & Using this code augmentation would create good habits when dealing with code. & \ref{CT-Habit}\\
                16 & After using this code augmentation, it would be hard to work without it. & \ref{CT-Habit}\\
            \hline
                17* & Has a positive attitude towards the functionality of code augmentation and a genuine interest in utilizing it to enhance code comprehension, review, and development. & Behavioral Intention & \\
            \bottomrule
            \multicolumn{4}{p{18cm}}{* - This observation was derived from the codification process of the qualitative analysis carried out on open questions. The aim is to determine whether there exists a behavioral intention to use, as indicated by positive attitudes expressed during the survey. To determine the degree on Likert's unidimensional scale, the median of the responses from the first 16 questions was calculated, which reflects the extent of their positive attitude towards usage.}
             
        \end{tabular}
        }
    \end{center}
\end{table}

The code snippets, with the respective code augmentation, were presented in the browser so that the participant did not need access to any tool. This is aligned with the study's goal, which is to evaluate the acceptance and applicability of a wide range of scenarios according to the participants' perceptions. Moreover, the tools developed aimed to evaluate the feasibility of the approach implementation in current IDEs and do not implement all the scenarios. For each scenario, the participant had access to the scenario description and code snippets that exemplified how it enhanced the code visualization.

The results of the UTAUT 2 survey were then analyzed using descriptive and inferential statistics to provide objective answers to the following research questions (RQ):

\begin{enumerate}[label=$RQ_{\arabic*}$, leftmargin=25mm, topsep=0pt,itemsep=-1ex,partopsep=1ex,parsep=1ex]
    \item-- What is the general acceptance of code augmentation?
    \label{RQ1}
    \item-- Which scenarios are more suitable for \textit{Code Review}, \textit{Refactoring,} and \textit{Team Efficiency}?  
    \label{RQ2}
    \item-- What are the rankings of scenario acceptance?    
    \label{RQ3}
    \item-- What are the primary findings from each individual scenario?
    \label{RQ4}
\end{enumerate}

\subsection{Ethical Implications}
Researchers must carefully consider the ethical implications of any procedures conducted in a survey involving human subjects or sensitive personal information. When risks surpass the minimum threshold, conducting a comprehensive and independent ethical evaluation before beginning any survey activities is essential. The researchers in this study adhered to the guidelines of the participating institutes and universities and concluded that the potential risks were minimal, leading to the study not being evaluated by an Institutional Review Board.

To ensure the protection of participants, the necessary measures were taken to maintain anonymity, safeguard data, and uphold confidentiality. Informed consent was obtained from all participants, focusing on their voluntary and anonymous involvement. Additionally, the study was designed to pose no potential harm to participants.

\subsection{Dissemination}
The survey was conducted remotely using Google Forms, with the aim of gathering responses from individuals with professional or academic backgrounds. Participants were invited to participate in the questionnaire and were encouraged to share the link to the survey within their respective academic and professional networks. The survey was distributed to software development experts in private companies and academic institutions. 

\subsection{Evaluation Method}

\subsubsection{Quantitative Evaluation}
\begin{figure}[!htb]
    \centering
    \includegraphics[width=10cm]{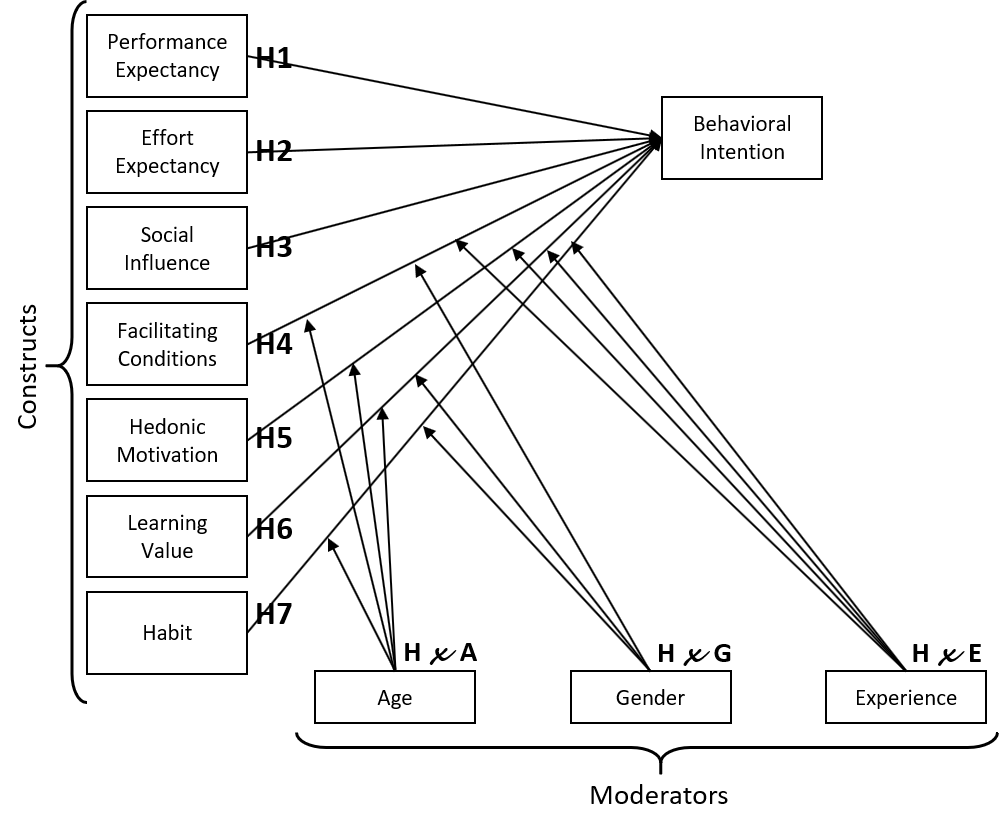}
    \caption{UTAUT 2 Complete Model Hypothesis} 
    \label{fig:UtautHip}
\end{figure}

In order to understand the behavior of system programmers, this study aimed to analyze the main characteristics that influence their intention to use code augmentation technology during the system development process. For this analysis, the UTAUT 2 model was adapted to the use of technology, performing a quantitative evaluation of the survey results using code augmentation. By implementing this methodology, still considering the survey instrument specified in the survey design as described in Section~\ref{surveydesignsection}, it is possible to establish the following hypotheses, according to the factors of the UTAUT 2 model (Figure~\ref{fig:UtautHip}):

\begin{enumerate}[label={$H_{\arabic*}$}, leftmargin=20mm, topsep=0pt,itemsep=-1ex,partopsep=1ex,parsep=1ex]
    \item-- The performance expectation positively influences the intention to use code augmentation.
    \label{h1}
    \item-- The expectation of effort positively influences the intention to use code augmentation.
    \label{h2}
    \item-- Social influence positively influences the intention to use code augmentation.
    \label{h3}
    \item-- The facilitating conditions positively influence the intention to use code augmentation.    
    \label{h4}
    \begin{enumerate}[leftmargin=10mm]
        \item[$H_{4}A_{1}$]-- The facilitating conditions are moderated by age group.
        \label{h4a}
        \item[$H_{4}G_{1}$]-- The facilitating conditions are moderated by gender.
        \label{h4g}
        \item[$H_{4}E_{1}$]-- The facilitating conditions are moderated by experience.
        \label{h4e}
    \end{enumerate}    
    \item-- Hedonic motivation positively influences the intention to use code augmentation.
    \label{h5}
    \begin{enumerate}[leftmargin=10mm]
        \item[$H_{5}A_{1}$]-- The Hedonic motivation are moderated by age group.        
        \label{h5a}
        \item[$H_{5}E_{1}$]-- The Hedonic motivation are moderated by experience.
        \label{h5e}
    \end{enumerate}    
    \item-- Learning positively influences the intention to use code augmentation.
    \label{h6}
    \begin{enumerate}[leftmargin=10mm]
        \item[$H_{6}A_{1}$]-- The value they place on education is moderated by age group.
        \label{h6a}
        \item[$H_{6}G_{1}$]-- The value they place on education is moderated by gender.
        \label{h6g}
        \item[$H_{6}E_{1}$]-- The value they place on education is moderated by experience.
        \label{h6e}
    \end{enumerate}    
    \item-- Habit positively influences the intention to use code augmentation.
    \label{h7}
    \begin{enumerate}[leftmargin=10mm]
        \item[$H_{7}A_{1}$]-- The habit are moderated by age group.
        \label{h7a}
        \item[$H_{7}G_{1}$]-- The habit are moderated by gender.
        \label{h7g}
        \item[$H_{7}E_{1}$]-- The habit are moderated by experience.
        \label{h7e}
    \end{enumerate}    
\end{enumerate}

The model was evaluated by checking Cronbach's Alpha (A.C.) \cite{Bland1997}, internal consistency using Composite Reliability (C.R.), convergent validity analyzed using Average Variance Extracted (A.V.E.), and discriminant validity estimated using the cross-loading criterion \citep{Henseler2009}.

As suggested by \citet{Venkatesh2012}, the UTAUT is employed to identify the pivotal factors affecting an individual's or group's willingness to adopt a new technology. It is not a hypothesis test in the statistical sense; rather, it is a conceptual model crafted to comprehend the underlying factors that influence technology adoption. Consequently, it does not incorporate the traditional null and alternative hypotheses typically associated with statistical tests such as the Student's t-test or the q-square test. Instead, the UTAUT focuses on elucidating the theoretical relationships among the mentioned constructs to explain technology adoption behavior.

In empirical research, null and alternative hypotheses are used to test statistical relationships or differences between variables. This involves formulating a null hypothesis, which typically states that there is no relationship or difference between the variables, and an alternative hypothesis, which asserts that there is a significant relationship or difference. These hypotheses are then tested using statistical methods \cite{Venkatesh2003, Venkatesh2012}.

In the case of UTAUT, it is a theoretical framework that provides a conceptual model for understanding why people adopt or do not adopt technology. UTAUT is not an experiment in itself and does not directly engage in statistical hypothesis testing. Instead, it provides a conceptual framework that researchers can use to formulate their own hypotheses based on the specific variables they are studying and the relationships they wish to investigate \cite{Venkatesh2003, Venkatesh2012}.

As defined by \citet{Venkatesh2003}, in the UTAUT Model, it's important to emphasize that hypotheses are formulated to assess the relationship between constructs, with the goal of formalizing clear expectations regarding the interconnections among variables in the model. Each hypothesis represents a verifiable assumption concerning how these variables are linked, making it easier to conduct specific empirical studies to support or challenge these relationships. This not only improves the testability and efficiency of research but also contributes to the advancement of knowledge in the field of technology acceptance and utilization, as empirical findings either confirm or refute theoretical predictions \cite{Venkatesh2003, Venkatesh2012}.

For each of the eleven (11) constructed scenarios, the UTAUT 2 model was applied and evaluated in isolation to respond to the defined hypotheses. The collected data was organized and descriptively analyzed using the SmartPLS\footnote{\url{https://www.smartpls.com}} software. In order to analyze the most relevant characteristics of the intention to use code augmentation through the studied variables, the study was carried out through the analysis of structural equations.

\subsubsection{Qualitative Evaluation}

In this research, a coding methodology was employed to conduct a comprehensive qualitative assessment, specifically focusing on analyzing the open-ended and free-response segments of the survey. This methodology entails a systematic process of assigning codes to individual responses, which involves a sequential examination of each line of gathered data \citep{thomas2008methods}. 

\begin{figure}[!ht]
    \centering
    \includegraphics[width=13cm]{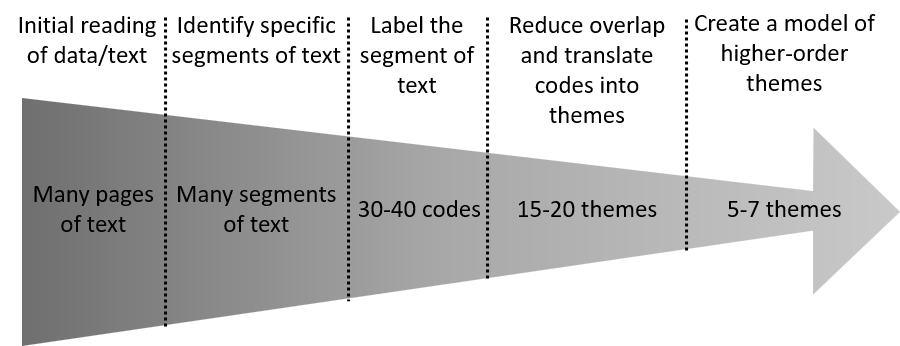}
    \caption{Coding process for qualitative analysis (adapted from \citep{cresswell2012}).}
    \label{fig:qualitativeeval}
\end{figure}

The inductive approach allows for the emergence of codes and themes organically from the collected responses (as depicted in Figure~\ref{fig:qualitativeeval}). In the context of software engineering, it is advisable to execute a series of steps, including \textit{data extraction}, \textit{coding of data}, and \textit{transformation of codes into thematic patterns} \citep{cresswell2012}). The systematic process improves our understanding of qualitative insights and facilitates the identification of underlying patterns and recurring themes that may have not yet been predefined.

\subsection{Threats to Validity}

According to \cite{wohlin2012}, this subsection discusses potential threats to the survey in terms of internal validity, external validity, construct validity, and conclusion validity.

\begin{enumerate}
    \item \textbf{Internal Validity:} Threats to internal validity concern the observed causal relationship between treatment and outcome that is not a result of the influence of another factor and cannot be controlled or measured \cite{wohlin2012}. To mitigate the threat, an approach was adopted that included the joint application of the UTAUT2 model and Structural Equation Modeling (SEM), along with the use of standardized instruments for measuring UTAUT2 variables, the implementation of appropriate statistical analyses considering the UTAUT2 model structure, and continuous monitoring of the model throughout the process.
    \item \textbf{External Validity:} The threats to external validity concern the generalization of results to industrial practice \cite{wohlin2012}. This survey was conducted within the context of evaluating the acceptance of identifiers and represents the initial phase of a broader, real-world industrial experiment. Ideally, future studies should involve the construction of practical experiments and replication on a larger scale within an industrial environment. However, all participants involved have extensive experience and are industry professionals, which may somewhat mitigate this threat in the context of this research.
    \item \textbf{Construct Validity:} Analyses that consider the relationships between theory and observation, that is, whether the treatment (method) accurately reflects the cause and whether the outcome accurately reflects the effect, represent threats to construct validity \cite{wohlin2012}. To mitigate this threat, the UTAUT2 model was used in conjunction with SEM and descriptive and inferential statistical analyses, such as bootstrapping, which ensure good reliability of the analyses.
    \item \textbf{Conclusion Validity:} Threats to conclusion validity are related to issues that may affect our analysis of the experiment's results, such as sample size selection, statistical test selection, and caution in implementing and analyzing an experiment \cite{wohlin2012}. The number of participants is too low to detect a significant effect in this experiment. The study's validity is enhanced by the use of highly qualified participants; however, the data should be viewed as indicative, not conclusive. However, to mitigate this threat and add greater reliability to the survey analysis, the bootstraping technique was employed. By applying bootstraping, more robust estimates of parameters and statistics of interest were obtained, thus reducing the risk of erroneous conclusions based on limited or biased samples. The benefits of bootstraping include the ability to construct confidence intervals and estimate variance and standard error more accurately, as well as to perform robust hypothesis tests that do not rely on specific assumptions about the data distribution.
\end{enumerate}

The UTAUT 2 model is a widely recognized framework for understanding technology adoption and use, especially in the context of information systems and technology. However, when applying UTAUT 2 in the context of Structural Equation Modeling (SEM) \cite{Jaehyeon2019}, several threats can emerge that affect the validity and reliability of the results \cite{Venkatesh2012}.  Some of the main threats include the following:

\begin{enumerate}
       \item \textbf{Context Specificity}: UTAUT 2 was developed based on empirical studies across different contexts. Specific context characteristics can influence the model's relationships between constructs and variables. Therefore, generalizing the results to a specific context may be a threat if the constructs do not behave similarly.
       \begin{itemize}
           \item \textit{\textbf{Mitigation Action:} To mitigate this threat, it is essential to conduct a contextual validation of the model. In response to this concern, several trials were conducted with a generic dataset on the proposed model to ensure its suitability before being applied to the specific case study data examined in this work.}
        \end{itemize}
        
       \item \textbf{Construct Measurement:} Proper operationalization and measurement of constructs are crucial in SEM. Any imprecision in measuring latent variables can affect the validity of the results. If measures are not reliable or valid, the model's outcomes may not accurately reflect the expected theoretical relationships.
       \begin{itemize}
           \item \textit{\textbf{Mitigation Action:} To address this threat, meticulous attention must be given to construct measurement, utilizing established measurement scales, validation procedures, and sensitivity analyses to ensure the reliability and validity of measures in SEM analysis. Model evaluation involved examining Cronbach's Alpha (CA) for internal consistency, followed by Composite Reliability (CR). Convergent validity was assessed using Average Variance Extracted (AVE), and discriminant validity was determined using the cross-loadings criterion.}
        \end{itemize}

       \item \textbf{Sampling:} The sample quality used for SEM analysis is pivotal. A non-representative or inadequately sized sample can lead to distorted or invalid results. Sampling should be carefully planned to avoid biases and ensure the representativeness of the target population.
       \begin{itemize}
           \item \textit{\textbf{Mitigation Action:} To mitigate this threat, it is essential to pay careful attention to the quality of sampling. Using a representative and adequately sized sample is crucial for SEM analysis. Meticulously planning the sampling process helps to avoid biases and ensures the representativeness of the target population. In this study, the bootstraping technique was employed, which can be used to assess the robustness of the results. By utilizing bootstrapping in SEM, we can generate multiple data samples from the original sample, allowing us to assess the variability of the estimated parameters and calculate confidence intervals for these parameters \citep{hair2014}. This helps to mitigate potential biases and increase the reliability of the conclusions drawn in SEM analysis.}
        \end{itemize}
        
       \item \textbf{Endogeneity:} Causal relationships between constructs can be affected by endogeneity, where variables are interrelated in ways not considered in the model. This can result in spurious effects and distort parameter estimates \citep{Henseler2009, Bland1997}.
       \begin{itemize}
           \item \textit{\textbf{Mitigation Action:} To mitigate endogeneity, the discriminant validity technique was used in this study. Discriminant validity helps verify if the model constructs are distinct from each other, minimizing the overlap of latent variables and thus contributing to a clearer interpretation of the specific effects of each construct. By ensuring that the constructs are truly different, discriminant validity reduces the likelihood of unconsidered interrelationships in the model, thereby mitigating endogeneity.}
        \end{itemize}
        
       \item \textbf{Collinearity:} The presence of collinearity among latent variables can complicate result interpretation. Collinearity may arise when variables are highly correlated, making it difficult to identify which variables contribute independently to construct an explanation \citep{Henseler2009, Bland1997}.
       \begin{itemize}
           \item \textit{\textbf{Mitigation Action:} To mitigate the threat of collinearity, this study conducted a diagnostic analysis of multicollinearity to identify highly correlated variables and their impact on the model by calculating variance inflation factors (VIFs) to assess the severity of collinearity.}
        \end{itemize}

       \item \textbf{Respondent Bias:} How participants respond to questionnaires can introduce bias into collected data. This can occur due to socially desirable responses, lack of question comprehension, or even memory errors.
       \begin{itemize}
           \item \textit{\textbf{Mitigation Action:} To mitigate the threat of respondent bias, this study implemented a careful questionnaire design to minimize ambiguity and ensure clarity of the questions. Additionally, instructions were developed clearly, and standardized scales were used to help improve participant understanding and reduce response errors.}
        \end{itemize}
        
\end{enumerate}

Hence, when applying the UTAUT 2 model within SEM, addressing these threats is important through careful research planning, rigorous data collection, meticulous analysis, and appropriate interpretation of findings. Additionally, acknowledging limitations and potential threats to validity when reporting results is crucial for robust and reliable analysis.

To mitigate threats when applying the UTAUT 2 model in SEM, it is essential to establish strong theoretical validation of proposed relationships, employ appropriate sampling, and ensure data cleanliness. 

Testing SEM assumptions, ensuring measures' convergent and discriminant validity, and addressing endogeneity are crucial steps. Additionally, selecting an appropriate model, conducting cross-validation, sensitivity analyses, and seeking input from expert reviewers contribute to more reliable outcomes. Interpretation of results should be cautious, accounting for limitations and potential biases. Replicating the analysis across different datasets enhances the robustness of the result.

 \section{Survey Results}
\label{results}

This section provides a context for the in-depth analysis of the empirical findings of this study. All the data used in this research can be obtained through the replication package available at the Zenodo.org\footnote{\url{https://zenodo.org/records/10888469}} repository. The analysis and interpretation of empirical evidence were carried out with a focus on the validity and reliability of the investigated constructs. The objective was to gain a comprehensive understanding of the underlying relationships and their theoretical and practical implications. Additional information regarding the process of analyzing reliability (Cronbach's Alpha, Composite Reliability, and Average Variance Extracted) and the survey results can be found in Appendix \ref{SurveyAnalysis} for a more comprehensive understanding.

The assessment of measurement reliability is a fundamental aspect of result interpretation. A reliability test in scientific research is a measure used to assess the consistency and stability of responses or measurements obtained from a data collection instrument (such as a questionnaire, scale, or test) over time or in different situations. Reliability is an important indicator of data quality and measurement consistency that allows for a degree of confidence in extrapolating and generalizing results beyond the tested population group.

The overall observation of Cronbach's Alpha values consistently equal to or higher than 0.6 suggests that each construct's items are internally cohesive, denoting a reliable measurement of the constructs in question.

Composite Reliability scores, employed to assess the internal consistency of constructs, consistently exceed the recommended threshold of 0.7. This trend reinforces the reliability of the measurement, indicating that the constructs are solidly capturing the underlying variability.

Average Variance Extracted (AVE) values provided insights into the extent to which the variance captured by construct items surpasses variance due to measurement error. In many scenarios, AVE values surpass the acceptable threshold of 0.5, signaling satisfactory convergent validity. This suggests that the constructs adequately represent the variability they aim to capture. The coherence and reliability observed in Composite Reliability and AVE scores across various scenarios enhance the robustness of the measurement model. The uniformity of reliability measures across constructs and scenarios suggests that questionnaire items consistently measure intended constructs with high precision and consistency.

In the end, these reliability metrics strengthen the robustness of the measurement model and engender confidence in the data collection procedure, thereby facilitating subsequent analyses and interpretations. In addition, a comprehensive examination of discriminant validity was performed, analyzing the interrelationships between constructs. The investigation of discriminant validity contributes to a greater understanding of inter-concept relationships and provides a comprehensive view of emerging implications for theory and practice.

It is essential to emphasize that, as predicted by \citet{Venkatesh2003}, hypotheses are created to assess the relationship between constructs, meaning that one hypothesis is formulated for each predicted relationship.
The analysis of the proposed hypotheses yielded significant insights into the relationships between the investigated constructs and their influence on behavioral intention. Eleven distinct scenarios were examined using the bootstrapping method in conjunction with the Student's t-test in order to examine these relationships, and the hypotheses were evaluated as follows:

\begin{itemize}[topsep=0pt,itemsep=-1ex,partopsep=1ex,parsep=1ex]
    \item \textbf{Hypothesis \ref{h1}}, which posited a positive relationship between \textbf{Performance Expectancy} and \textbf{Behavioral Intention}, was confirmed in scenario 11 (\ref{sc11}-Person updated). The result was supported by a t-value of 3.467, a p-value of 0.001, and an R-Square ($R^2$) of 0.652, indicating a substantial explanatory power of 65.2\%.
    
    \item \textbf{Hypothesis \ref{h2}}, suggesting an association between \textbf{Effort Expectancy} and \textbf{Behavioral Intention}, was corroborated in scenarios 01 (\ref{sc1}-Descriptive names), 02 (\ref{sc2}-Abbreviated names), 06 (\ref{sc6}-Showing modifiers), 07 (\ref{sc7}-Showing annotation), 09 (\ref{sc9}-Problems detected), 10 (\ref{sc10}-Visual reminders), and 11 (\ref{sc11}-Person updated). t-values ranged between 2.01 and 2.707, p-values ranged from 0.007 to 0.045, and R-Squares ($R^2$) ranged from 0.547 to 0.652. These findings consistently demonstrate that a favorable perception of effort contributes to a higher behavioral intention.
    
    \item \textbf{Hypothesis \ref{h3}}, exploring the influence of \textbf{Social Influence} on \textbf{Behavioral Intention}, was validated in scenario 05 (\ref{sc5}-New update). With a t-value of 2.009, a p-value of 0.045, and an R-Square ($R^2$) of 0.622, a moderate effect of \textbf{Social Influence} on \textbf{Behavioral Intention} was evident.

    \item Conversely, \textbf{Hypothesis \ref{h4}}, which sought a relationship between \textbf{Facilitating Conditions} and \textbf{Behavioral Intention}, found no support in any of the analyzed scenarios.

    \item Regarding \textbf{Hypothesis \ref{h5}}, exploring the relationship between \textbf{Hedonic Motivation} and \textbf{Behavioral Intention}, none of the scenarios provided support for this association.
    
    \item \textbf{Hypothesis \ref{h6}}, concerning the association between \textbf{Learning Value} and \textbf{Behavioral Intention}, was corroborated in scenarios 03 (\ref{sc3}-Bad naming) and 04 (\ref{sc4}-Recent renames). t-values ranged from 2.318 to 2.723, p-values ranged from 0.006 to 0.02, and R-Squares ($R^2$) ranged from 0.616 to 0.738. This indicates that learning value positively influences behavioral intention.
    
    \item \textbf{Hypothesis \ref{h7}}, addressing the association between \textbf{Habit} and \textbf{Behavioral Intention}, was confirmed in scenario 06 (\ref{sc6}-Showing modifiers), with a t-value of 2.43, a p-value of 0.015, and an R-Square ($R^2$) of 0.6, underscoring the relevant role of habit in predicting behavioral intention.
\end{itemize}

The investigated moderators did not demonstrate a consistent moderating effect on the relationships between constructs and behavioral intention in the examined scenarios.
In the context of code augmentation, these analyses highlight the significance of specific constructs such as \textbf{Performance Expectancy}, \textbf{Effort Expectancy}, \textbf{Learning Value}, and \textbf{Social Influence}. A nuanced comprehension of these relationships contributes to a broader understanding of the determinants of user behavior in this particular context, thereby providing valuable insights for theory and practice.

The analysis of various scenarios and hypotheses sheds light on the factors influencing the intention to use code augmentation. The findings reveal distinct patterns of influence, allowing a comprehensive understanding of the interplay between different constructs and their impact on user behavior.

\subsection{Statistical results}

Based on a combination of the results obtained from statistical analyses conducted for each hypothesis across different scenarios, it is possible to determine the ranking and classification of the influence of scenarios on the use of code augmentation. t-values indicate the size of the statistical effect, while p-values indicate statistical significance. A higher magnitude t-value suggests a stronger influence and a lower p-value indicates greater statistical confidence in the relationship. R-Square ($R^2$) measures the proportion of variability in the response variable explained by the model.

Combining these indices for each hypothesis and scenario makes it possible to classify and rank the influence of variables on the intention to use code augmentation. Hypotheses and scenarios with higher t-values, lower p-values, and higher R-Squares ($R^2$) tend to have a stronger and more significant influence on usage intention. This enables the identification of the most important variables and their corresponding scenarios of greater relevance in explaining the adoption behavior of code augmentation.
Table~\ref{table:ScenarioRaking} presents the results of hypothesis analysis in different scenarios along with relevant statistical metrics.

\begin{table}[width=.9\linewidth,pos=h]
  \begin{center}	
\caption{Ranking of hypotheses and scenarios based on statistical metrics.}
  \label{table:ScenarioRaking}
\resizebox{16.4cm}{!}{
  \begin{tabular}{c*{2}{l}*{4}{c}}
		  \toprule \textbf{Ranking} & \textbf{Hypothesis} & \textbf{Scenario} & \textbf{t-value} & \textbf{p-value} & \textbf{$R^2$} & \textbf{Rating} \\
            \midrule
                01 & \ref{h1}-Performance Expectancy & (\ref{sc11}) Person updated & 3.467 & 0.001 & 0.652 & High \\
                02 & \ref{h2}-Effort Expectancy & (\ref{sc11}) Person updated & 2.111 & 0.035 & 0.652 & High \\
            \hline
                03 & \ref{h6}-Learning Value & (\ref{sc3}) Bad naming     & 2.723 & 0.006 & 0.738 & Moderate \\
                04 & \ref{h3}-Social Influence & (\ref{sc5}) New update     & 2.009 & 0.045 & 0.622 & Moderate \\
                05 & \ref{h6}-Learning Value & (\ref{sc4}) Recent renames & 2.318 & 0.020 & 0.616 & Moderate \\
                06 & \ref{h7}-Habit & (\ref{sc6}) Showing modifiers & 2.430 & 0.015 & 0.600 & Moderate \\
            \hline
                07 & \ref{h2}-Effort Expectancy & (\ref{sc1}) Descriptive names  & 2.705 & 0.007 & 0.251 & Low \\
                08 & \ref{h2}-Effort Expectancy & (\ref{sc7}) Showing annotation & 2.090 & 0.037 & 0.546 & Low \\
                09 & \ref{h2}-Effort Expectancy & (\ref{sc2}) Abbreviated names  & 2.442 & 0.015 & 0.547 & Low \\
                10 & \ref{h2}-Effort Expectancy & (\ref{sc9}) Problems detected  & 2.052 & 0.040 & 0.584 & Low \\
                11 & \ref{h2}-Effort Expectancy & (\ref{sc10}) Visual reminders  & 2.707 & 0.007 & 0.612 & Low \\
            \hline
                - & \ref{h4}-Facilitating Conditions & - & - & - & - & Not Confirmed \\
                - & \ref{h5}-Hedonic Motivation & - & - & - & - & Not Confirmed \\
                - & Moderators & - & - & - & - & Not Confirmed \\                
		  \bottomrule
                & & & & \multicolumn{1}{l}{$R^2$ Mean}   & 0.585 \\
                & & & & \multicolumn{1}{l}{$R^2$ Median} & 0.606 \\
                & & & & \multicolumn{1}{l}{$R^2$ StDev}  & 0.117 \\
    \end{tabular}
    }
    \end{center}
\end{table}

Although the hypotheses \ref{h1} and \ref{h2} have a lower R-Square ($R^2$) compared to \ref{h6}, they are ranked higher on the list due to consideration of factors beyond R-Square ($R^2$) in determining the ranking. The key determinant for this classification was that hypotheses \ref{h1} and \ref{h2} could be validated by the same scenario.

The classification of hypotheses regarding the use of code augmentation is conducted based on R-Square ($R^2$), which indicates the proportion of variability in usage intention explained by relationships between constructs. The higher the R-Square ($R^2$), the stronger the influence of the independent construct (predictor variable) on the dependent construct (usage intention). The classification is divided into three categories:

\begin{enumerate}[topsep=0pt,itemsep=-1ex,partopsep=1ex,parsep=1ex]
    \item High: Hypotheses with a significantly high R-Square ($R^2$), indicating that the relationship between constructs has a substantial influence on the intention of use. This suggests that variations in the independent constructs explain a large portion of the variation in usage intention.
    \item Moderate: Hypotheses with a moderate R-Square ($R^2$), indicating that the relationship between constructs has a reasonable influence on usage intention. While not as strong as in the ``High'' category, there is still a significant influence.
    \item Low: Hypotheses with a relatively low R-Square ($R^2$), indicating that the relationship between constructs has a limited influence on usage intention. In this case, variations in independent constructs contribute less to explaining the variation in usage intention.
\end{enumerate}

Unconfirmed hypotheses are marked as ``Not Confirmed'' and are not classified, as they did not provide sufficient statistical evidence to demonstrate a significant relationship between constructs.

Moderators were also not confirmed and are classified as ``Not Confirmed'', indicating that there is not enough evidence to show that they influence the relationship between constructs in the tested hypotheses.

In summary, the ranking provides insight into the relative influence of each hypothesis and scenario on code augmentation usage, based on a statistical assessment of relationships between constructs. This helps identify factors with a greater or lesser impact on the intention to use this type of technology. The main findings and insights are:

\begin{enumerate}[topsep=0pt,itemsep=-1ex,partopsep=1ex,parsep=1ex]
    \item \textbf{High Influence Scenarios}
    \begin{enumerate}[topsep=0pt,itemsep=-1ex,partopsep=1ex,parsep=1ex]
        \item \textbf{Scenario 11 (\ref{sc11}):} Showing the last person that updated some code;
        \begin{enumerate}[topsep=0pt,itemsep=-1ex,partopsep=1ex,parsep=1ex]
\item \textbf{Findings:} In this scenario, both performance expectancy and effort expectancy significantly contribute to users' intention to use code augmentation. Users perceive that the technology enhances their performance and is easy to use, thereby increasing their willingness to adopt it.
        \end{enumerate}
    \end{enumerate}
    \item \textbf{Moderate Influence Scenarios}
    \begin{enumerate}[topsep=0pt,itemsep=-1ex,partopsep=1ex,parsep=1ex]
        \item \textbf{Scenario 03 (\ref{sc3}):} Variables with bad naming practices;
        \begin{enumerate}[topsep=0pt,itemsep=-1ex,partopsep=1ex,parsep=1ex]
\item \textbf{Findings:} Learning value has a moderate effect on users' intention to use code augmentation. Users perceive that the technology offers valuable learning opportunities, leading to increased adoption.
        \end{enumerate}
        \item \textbf{Scenario 05 (\ref{sc5}):} New methods and updated methods;
        \begin{enumerate}[topsep=0pt,itemsep=-1ex,partopsep=1ex,parsep=1ex]
\item \textbf{Findings:} Social influence moderately impacts users' intention to use code augmentation. The opinions and behaviors of peers or colleagues influence users' decisions to adopt the technology.
        \end{enumerate}        
        \item \textbf{Scenario 04 (\ref{sc4}):} Recent renames;
        \begin{enumerate}[topsep=0pt,itemsep=-1ex,partopsep=1ex,parsep=1ex]
\item \textbf{Findings:} Similar to Scenario 3, learning value moderately influences users' intention to adopt code augmentation. The perception of meaningful learning outcomes drives users' willingness to use the technology.
        \end{enumerate}
        \item \textbf{Scenario 06 (\ref{sc6}):} Showing the presence of modifiers in method usage;
        \begin{enumerate}[topsep=0pt,itemsep=-1ex,partopsep=1ex,parsep=1ex]
\item \textbf{Findings:} In this scenario, both ease of use (effort expectancy) and habitual behavior significantly impact users' intention to use code augmentation. Habitual usage and perceived ease of use drive adoption.
        \end{enumerate}
    \end{enumerate}
    \item \textbf{Low Influence Scenarios:}
    \begin{enumerate}[topsep=0pt,itemsep=-1ex,partopsep=1ex,parsep=1ex]
        \item \textbf{Scenario 01 (\ref{sc1}):} Showing more descriptive names;
        \begin{enumerate}[topsep=0pt,itemsep=-1ex,partopsep=1ex,parsep=1ex]
\item \textbf{Findings:} While ease of use is a contributing factor, it holds a lower influence compared to other scenarios. Users' intention to use code augmentation is shaped by the perceived effort required.
        \end{enumerate}
        \item \textbf{Scenario 07 (\ref{sc7}):} Showing the presence of annotation on method and class usage;
        \begin{enumerate}[topsep=0pt,itemsep=-1ex,partopsep=1ex,parsep=1ex]
\item \textbf{Findings:} Ease of use plays a minor role in influencing users' intention to use code augmentation in this scenario.
        \end{enumerate}
        \item \textbf{Scenario 02 (\ref{sc2}):} Showing abbreviated names;
        \begin{enumerate}[topsep=0pt,itemsep=-1ex,partopsep=1ex,parsep=1ex]
\item \textbf{Findings:} Similar to Scenario 1, ease of use has a relatively lower impact on users' intention to adopt code augmentation.
        \end{enumerate}        
        \item \textbf{Scenario 09 (\ref{sc9}):} Problems detected through static analysis tools;
        \begin{enumerate}[topsep=0pt,itemsep=-1ex,partopsep=1ex,parsep=1ex]
\item \textbf{Findings:} Similar to previous scenarios, perceived ease of use has a limited influence on users' intention to adopt code augmentation.
        \end{enumerate}
        \item \textbf{Scenario 10 (\ref{sc10}):} Visual reminders on API usage;
        \begin{enumerate}[topsep=0pt,itemsep=-1ex,partopsep=1ex,parsep=1ex]
\item \textbf{Findings:} In this scenario, as in others, ease of use moderately contributes to users' intention to use code augmentation.
        \end{enumerate}
    \end{enumerate}
    \item \textbf{Not confirmed Hypotheses:}
    \begin{enumerate} [topsep=0pt,itemsep=-1ex,partopsep=1ex,parsep=1ex]        
        \item \textbf{Hypothesis \ref{h4} (Facilitating Conditions $\rightarrow$ BI):} Not confirmed.
        \begin{enumerate}[topsep=0pt,itemsep=-1ex,partopsep=1ex,parsep=1ex]
            \item \textbf{Findings:} Facilitating conditions do not play a significant role in shaping users' intention to use code augmentation in this context.
        \end{enumerate}
        \item \textbf{Hypothesis \ref{h5} (Hedonic Motivation $\rightarrow$ BI):} Not confirmed.
        \begin{enumerate} [topsep=0pt,itemsep=-1ex,partopsep=1ex,parsep=1ex]          
            \item \textbf{Findings:} Hedonic motivation has negligible influence on users' intention to adopt code augmentation.
        \end{enumerate}
    \end{enumerate}
\end{enumerate}

In conclusion, scenarios and hypotheses have varying degrees of influence on the intention of users to utilize code augmentation. \textbf{Performance expectancy}, \textbf{effort expectancy}, \textbf{learning value}, and \textbf{social influence} emerge as key motivators, whereas \textbf{facilitating conditions} and \textbf{hedonic motivation} have limited or no influence.

The R-Square ($R^2$) Mean of 0.585 indicates that the relationships between constructs explain approximately 58.5\% of the variation in usage intention. The higher R-Square ($R^2$) Median (0.606) suggests that many scenarios exhibit a strong relationship between the constructs and the Behavioral Intention. Remember that the R-Square ($R^2$) indicates the proportion of variability in the response variable that is accounted for by the model, and higher R-Squares ($R^2$) generally correspond to a more pronounced and substantial impact on usage intention. In other words, a higher R-Square ($R^2$) value means that there is a stronger relationship between the constructs of the model and Behavioral Intention.

Combining these indices for each hypothesis and scenario makes it possible to classify and rank the influence of variables on the intention to use code augmentation. Hypotheses and scenarios with higher t-values, lower p-values, and higher R-Squares ($R^2$) tend to have a stronger and more significant influence on usage intention. This enables the identification of the most important variables and their corresponding scenarios of greater relevance in explaining the adoption behavior of code augmentation.
Table~\ref{table:ScenarioRaking} presents the results of hypothesis analysis in different scenarios along with relevant statistical metrics.

In conclusion, the R-Square ($R^2$) statistics collectively suggest that the constructed model adequately captures the variance in the intention to use variables across various scenarios. Although individual scenarios may exhibit varying levels of influence, the model's moderate average R-Square ($R^2$), higher median, and low dispersion indicate a reasonable and consistent level of explanatory ability. This bolsters the confidence in the model's reliability and its potential to provide meaningful insights into the factors influencing the use of code augmentation.

\subsection{Thematic Analysis}

Thematic analysis is a commonly used technique in social, psychological, educational, and other humanities areas to analyze qualitative data. This approach is particularly useful for identifying, analyzing, and interpreting patterns of meaning (themes) within a dataset.

In the process of thematic analysis, the content of questionnaire responses was carefully examined to identify recurring patterns, common topics, concepts, or emerging ideas in the data.

Following the steps outlined by Figure \ref{fig:qualitativeeval}, the process proceeded objectively as follows:

\begin{enumerate}
    \item \textbf{Familiarization with the data:} Data readings were conducted multiple times to become familiar with the content and gain a general understanding;
    \item \textbf{Generation of codes:} Codes were assigned to relevant segments of the data reflecting key concepts, ideas, or themes related to transient identifiers and programming practices;
    \item \textbf{Identification of themes:} Codes were grouped into broader themes based on similarities and conceptual relationships.
    \item \textbf{Review and refinement of themes:} Themes were reviewed, refined, and adjusted as the analysis progressed, ensuring they adequately captured the essence of the data.
\end{enumerate}

It is important to note that at least two authors were more intensively involved in the process of thematic analysis and code generation; however, all authors reviewed the process and provided critiques until it was in accordance with all authors.

One of the most crucial processes in thematic analysis is the generation of codes, which serves as the foundation for the entire research. While there are several techniques available, in this study, we opted for the method of double coding, where two or more independent researchers coded the same datasets to identify emerging themes. To address discrepancies and ensure the reliability of the results, the following strategy was employed:

\begin{enumerate}
    \item \textbf{Open and collaborative discussion:} The authors engaged in open and collaborative discussions to understand the different perspectives and approaches regarding the data. This involved meetings aimed at reviewing and debating the emerging themes, as well as resolving any conflicts or divergences.
    \item \textbf{Data review and triangulation:} All authors reviewed and triangulated data from various sources to verify the consistency of emerging patterns and codes.
\end{enumerate}

In cases of disagreement, authors who disagreed on the interpretation of a particular dataset described their differing perspectives and sought to reach a consensus or compromise. In our study, there was only one discrepancy related to the Target Audience, where one author classified it as \textit{``Experience''}. Following a discussion, consensus was reached to use the term \textit{``Target Audience''}. The remaining generated codes were consensually refined and grouped, particularly when one author excessively detailed a code that, in subsequent revisions, was grouped into a broader theme.


It is important to highlight that, overall, the questionnaire was constructed based on the UTAUT2 model and its responses were on the Likert scale. Additionally, two open-ended questions were included at the end for participants to provide their impressions on the use of impermanent identifiers. Thus, thematic analysis was conducted in addition to UTAUT2 in the open-ended questions, making the application of Inter-Rater Reliability (IRR) not suitable.

\subsubsection{Data Extraction}

The initial phase of thematic analysis involved the execution of the data extraction procedure \citep{cresswell2012}. The study received responses from 114 different individuals, all anonymously labeled P1 to P114, of which 108 were considered valid and included in the analysis.

It is noteworthy that the study comprises two open-ended questions:
\begin{enumerate}[topsep=0pt,itemsep=-1ex,partopsep=1ex,parsep=1ex]
    \item What is your opinion about using this code augmentation in general? Is it useful? Is there any specific public that would benefit from it? Do you see any drawbacks or problems in its usage?
    \item Do you have an idea of any additional scenarios for code augmentation not covered by one of the scenarios? Or maybe any variation of these scenarios? 
\end{enumerate}

The survey gathered feedback from 68 respondents regarding Question 1 and 45 respondents regarding Question 2. It is noteworthy that there is a degree of overlap, as the total number of respondents who answered at least one of the two questions was 68.

It is important to highlight that the entire process of thematic analysis was conducted and reviewed by at least two of the researchers involved in this publication.

\subsubsection{Coding Data}

The responses were entered into a spreadsheet, where columns were allocated for questions Q1 through Q2 based on their sequential order in the survey. Rows were designated for participants, identified as P1 through P108. Subsequently, the data encoding procedure was initiated, encompassing a per-question approach across all respondents. In this process, one or more codes were assigned to each of the 108 responses.

To illustrate the process, see some remarkable examples of response text: from participant P10 - \textit{``Ones which show possible problems with code are especially useful''} adds the code \textbf{``Problem Identification Agility''}. From the participant P13 - \textit{``If it's not too much can be very useful to make code review way faster and allows you to focus more on coding''}. In this fragment, we can directly infer the code \textbf{``Enhanced Refactoring''}. And from the participant P68  – \textit{``I see it as applicable. But without the use of emojis, emojis bring a different meaning to each one who reads, this causes greater difficulty than traditional textual writing''}, adds the code \textbf{``Ambiguity''}. 

\subsubsection{Translate Codes Into Themes}

The process of translating involved categorizing identified codes as either \textit{strengths} or \textit{weaknesses} of code augmentation use and grouping them under relevant top-level themes. Table~\ref{table:thematictable} presents the outcome of the aforementioned translation and the reduction of overlap.

\begin{table}[width=.9\linewidth,pos=h]
  \begin{center}	
  \caption{Themes obtained from the thematic analysis.}
  \label{table:thematictable}
\begin{tabular}{*{2}{l}r}
		  \toprule & \textbf{Codes} & \textbf{Total} \\
            \midrule
                \multicolumn{3}{l}{\textbf{Theme: Developer Experience}}\\
            \hline
                \multirow{3}{*}{Target audience} & Beginners & 21 \\
                                           & Experienced   & 09 \\
                                           & Senior & 09 \\
                                           \cline{2-3}
                \multirow{2}{*}{Strengths} & Syntax and Design Innovation & 07 \\
                                           & Impact on Code Visualization and Reading & 28 \\
                                           \cline{2-3}
                \multirow{2}{*}{Weaknesses} & Excessive Sensory Stimulation & 20 \\
                                           & Ambiguity & 35 \\
            \hline
                \multicolumn{3}{l}{\textbf{Theme: Architecture and Design}}\\
            \hline
                \multirow{5}{*}{Strengths} & Enhanced Clarity and Readability & 28 \\
                                           & Improved Code Quality & 24 \\
                                           & Problem Identification Agility & 15 \\
                                            & Enhanced Refactoring & 11 \\
                                           & Vulnerabilities Identification & 09 \\
                                           \cline{2-3}
                Weaknesses & (Not applicable) & - \\
            \hline
                \multicolumn{3}{l}{\textbf{Theme: Learning Curve}}\\
            \hline
                Strengths & Enhanced Learning & 11 \\
                \cline{2-3}
                Weaknesses & Blocks learning & 05 \\
\bottomrule            
    \end{tabular}
\end{center}
\end{table}

The table groups and identifies strengths and weaknesses associated with different aspects of this approach. The numbers correspond to the total number of individuals expressing each specific perception.

In the strengths and weaknesses table ~\ref{table:thematictable}, it is important to recognize that classifying the target audience into Beginners, Experienced, and Senior may initially seem inappropriate for this analysis. After all, this categorization refers to the audience's level of experience with the product, service, or project in question, rather than directly to its strengths and weaknesses.

However, classifying the Target Audience is crucial for understanding the context in which strengths and weaknesses are assessed. Each segment of the audience has different needs, expectations, and abilities, which directly influence how they perceive and interact with what is being offered.

By understanding the level of experience of the target audience, we can tailor our strategies, resources, and approaches according to their specific needs. For example, the characteristics that are considered strengths for beginners may not be perceived in the same way by experienced or senior users.

Therefore, although the classification of the Target Audience may not be a strength or weakness in itself, it plays a fundamental role in identifying and understanding the variables that shape the strengths and weaknesses of a product, service, or project. It is a key element for developing effective, user-centered strategies.

Thematic analysis reveals that the perception of code augmentation is predominantly positive, with many participants highlighting its benefits in terms of clarity, readability, quality, and the agility of the development process. The applicability of code augmentation to different levels of developer experience is also notable, underscoring its adaptability. However, concerns about potential downsides, such as ambiguity and sensory overload, are also raised. These insights provide a comprehensive view of the pros and cons of code augmentation and can guide decision-making regarding its adoption and implementation in software engineering practices.

\subsubsection{Thematic Discussion}

Through a qualitative examination of the responses, a comprehensive understanding of the strengths, weaknesses, and overall sentiments surrounding code augmentation has been attained. 
\vspace{5mm}

\textit{\textbf{5.2.4.1. Strengths}}

One of the primary insights derived from the analysis is the acknowledgment of the usefulness and benefits of code augmentation across different audience. The results of the thematic analysis reveal a positive sentiment, indicating that developers at various audience, ranging from beginners to seniors, recognize the value of implementing code augmentation. Indeed, it was the \textbf{``beginners''} code who received the highest number of recommendations, with citations such as:
\begin{itemize}
    \item \textit{``No problems at all, it would greatly benefit a team of beginners in programming''}
    \item \textit{``I think there are some interesting use cases that can help specially new developers. \ldots''}
\end{itemize}

The feedback received indicates that developers recognize the potential for code augmentation to impact the comprehension of code and its visual presentation directly. This insight reflects the significance of clean and intelligible code as a fundamental aspect of software development and highlights the role that code augmentation can play in advancing these qualities. The \textbf{``Impact on Code Visualization and Reading''} code who received the highest number of recommendations, with citations such as:
\begin{itemize}
    \item \textit{`` \ldots code augmentation proves generally beneficial, as it enhances code visualization and provides additional information \ldots''}
    \item \textit{``I think it's really useful and would help a better reading of the code.''}
\end{itemize}

These comments underscore the potential usefulness and adaptability of code augmentation techniques, which can cater to developers with diverse levels of experience and expertise. Furthermore, the analysis underscores the importance of enhancing code clarity, readability, and quality through augmentation. In this instance, codes \textbf{``Enhanced Clarity and Readability''} and \textbf{``Improved Code Quality''} received the most recommendations, with mentions such as:
\begin{itemize}
    \item Enhanced Clarity and Readability:
    \begin{itemize}
        \item \textit{``\ldots code augmentation in Scenario 1 and Scenario 8 could be valuable for creating self-explanatory code, eliminating the need for line-by-line comments. \ldots''}
        \item \textit{``Including emojis instead of errors simplifies understanding and reinforces communication between teams. \ldots''}
    \end{itemize}
    \item Improved Code Quality:
    \begin{itemize}
        \item \textit{``Generating code variations aids bug fixing, enhances performance, and explores features, valuable for large, complex codebases with time-consuming manual reviews.''}
        \item \textit{``Utilizing emojis for code comprehension can be a strategic choice for programmers, accelerating issue resolution; however, excessive emoji use may result in confusion. \ldots''}
    \end{itemize}
\end{itemize}

\vspace{10mm}

\textit{\textbf{5.2.4.2. Weaknesses}}

However, the analysis also raises concerns regarding potential drawbacks. The concept of \textbf{``excessive sensory stimulation''} and the risk of \textbf{``ambiguity''} in augmented code underscore the importance of a balanced implementation, as indicated by mentions such as:

\begin{itemize}
    \item Excessive Sensory Stimulation:
    \begin{itemize}
        \item \textit{``\ldots if too many features are used at once, I imagine it can become a bit messy or too much.''}
        \item \textit{``I feel like having to many emoji in the code might hinder its readability (information overload). \ldots''}
\end{itemize}
    \item Ambiguity:
    \begin{itemize}
        \item \textit{``\ldots emojis in code augmentations may require an adjustment period due to varying individual interpretations.''}
        \item \textit{``Drawbacks may result from misunderstandings or emoji interpretation lapses, primarily the responsibility of the programmer.''}
    \end{itemize}
\end{itemize}

This finding serves as a cautionary reminder that while augmentation can enhance certain aspects, care must be taken to ensure that it does not compromise the code's understandability or impede the learning process, such as dealing with information overload. We highlight some possible remedies for this in Section \ref{ImpIden}. Regarding ambiguity, some of the respondents propose some solutions, such as adding a textual description of the augmentation when hovering on an emoji.

In addition, some respondents were not receptive to the general idea of using emojis as a mechanism to augment code, e.g., \textit{``I don't believe it's really useful. Maybe it's because I'm an `old school' programmer''}, or \textit{``codes with emoji are the most ridiculous thing I ever saw.''}.
\vspace{5mm}

\textit{\textbf{5.2.4.3 Proposed Scenarios}}

Besides the thematic analysis based on open-ended questions, participants were queried regarding potential scenarios in which code augmentation could prove significantly beneficial. The responses indicate a strong endorsement of the technique, and the mentions provided illustrate that participants recognize its applicability in a variety of other contexts. Such as:

\begin{itemize}
    \item \textbf{Visualizing and Enhancing Code Readability}
    \begin{itemize}
    \item \textit{``Maybe one augmentation showing features like a deprecated method being used for instance \ldots''}
    \item \textit{``Using external and internal parameters in Swift method declarations, like 'serveCoffee(for person: Person)', enhances code readability.''}
    \end{itemize}

    \item \textbf{Parameters}
    \begin{itemize}
        \item \textit{``\ldots but I really like when you can read the declaration of the method as a sentence - including the parameters''}
        \item \textit{``\ldots maybe emojis for the parameters to understand what type they are \ldots''}
    \end{itemize}
    
    \item \textbf{Syntax and Semantics Enhancements}
    \begin{itemize}
    \item \textit{``Show when a variable named with a singular term is used as a type that allows multiple values (Array, List, ...), and the other way around''}
    \item \textit{``For the scenario 1, I would prefer if the name of the parameter would be added beside the value rather than in the name: addTranslation(word: `buon', translation: `good') with `word': and `translation': in a different color \ldots''}
\end{itemize}
    
    \item \textbf{Code Analysis and Program Structures}
    \begin{itemize}
\item \textit{`` \ldots Yes, thread-safe code in Java \ldots''}
    \item \textit{`` \ldots Emojis that indicate polymorphism points \ldots''}
    \item \textit{`` \ldots Emojis that indicate composition, injection, and hotspots  \ldots''}
    \item \textit{``For RFC (remote calls), could have a remote control / As we have a lot of infrastructure as code and several applications running in the cloud, there could be a scenario that assists in this regard \ldots''}
    \end{itemize}
    
    \item \textbf{Testing and Code Coverage Analysis}
    \begin{itemize}
    \item \textit{``How about unit test coverage analysis? Alerts for uncovered or potentially buggy code snippets would be beneficial. Also, indicate test coverage for methods. \ldots''}
    \end{itemize}
    
    \item \textbf{Documentation and Coding Conventions}
    \begin{itemize}
    \item \textit{``An Emoji for all the methods that have not a comment \ldots''}
    \item \textit{``An emoji that alerts the programmer if a standard coding convention is not being respected \ldots''}
    \item \textit{``Could include the emoji in @author for the methods and classes Javadoc \ldots''}
    \end{itemize}
    
    \item \textbf{Code Tracking and History}
    \begin{itemize}
    \item \textit{``Maybe indicating how long was the last change in each method (e.g., long ago, some time ago, recently). Or which code has been refactored \ldots''}
    \end{itemize}
    
    \item \textbf{Improved Error Messages}
    \begin{itemize}
    \item \textit{``Using code augmentation only in error messages (to better communicate the error) or warnings. See the Rust error messages as an example - they use heavy code decoration to \ldots''}
    \end{itemize}
\end{itemize}

The prominence of positive sentiment surrounding the concept of code augmentation indicates its general acceptance and recognition of its usefulness. Developers seem to appreciate the potential of augmentation to expedite refactoring, identify problems, and improve the overall quality of the code. This enthusiasm suggests a positive receptiveness toward incorporating augmentation practices into software development workflows. Developers recognize that augmentation may be more suitable for certain types of projects, audiences, or development stages. This insight underscores the flexibility and potential customizability of code augmentation methodologies to cater to diverse requirements.

\subsection{Overall Survey Discussion}

This section presents the discussions and general findings obtained from the analysis of the responses collected in the study on code augmentation through descriptive and inferential statistics, including Structural Equation Modeling (SEM).

Considering the descriptive statistics of Section~\ref{survey_results_overview}, the scenarios analyzed reveal a general trend of agreement among the participants. The answers indicate that the majority of participants demonstrated agreement or neutrality regarding the proposed scenarios. Response averages consistently ranged between 3.1 and 3.6, suggesting that the opinions remained closely aligned and positively assessed.

The distribution of responses, with a standard deviation generally ranging from 1.0 to 1.2, demonstrates a certain level of variability in opinions, as is typical in surveys of this nature. However, it is worth noting that the responses did not exhibit excessive dispersion relative to the means, which further underscores the consistency of the evaluations.
Consistently observed medians, primarily at 3 (``Don't agree or disagree'') or 4 (``Agree''), underscore that a majority of participants leaned toward selecting these values. This pattern implies a certain level of consensus in participants' opinions, enhancing confidence in the outcomes.

Regarding the distribution between response categories, the prominence of the ``Agree'' is notable and reinforces the positive sentiment of the assessments. Moreover, the relatively lower proportions in extreme categories like ``Strongly Agree'' and ``Strongly disagree'' imply a sense of equilibrium and significance in the conveyed opinions

Considering an overview of responses across all scenarios, the ``Agree'' category continues to emerge as the most prevalent, further underscoring the positive tendency in participants' feedback. Initial descriptive analysis serves as a strong indicator to address the research question \ref{RQ1}.

\begin{quote}
    \textbf{\ref{RQ1} - What is the general acceptance of code augmentation?}
    
    \textit{\textbf{\underline{Answer:}} Considering the analysis of Section \ref{SurveyAnalysis} and an overview of responses across all scenarios, the `Agree' category stands out, suggesting a trend of positive acceptance with some cohesion in opinions and moderate variability. Thus, it can be concluded that the results indicate a favorable reception of the technology.}
\end{quote}

Building upon the analysis of descriptive statistics, we can evaluate in which scenarios the practice of code augmentation is more suitable, thus addressing the research question \ref{RQ2}.

\begin{quote}
    \textbf{\ref{RQ2} - Which scenarios are more suitable for Code Review, Refactoring, and Team Efficiency?}

    \textit{\textbf{\underline{Answer:}} Among the analyzed scenarios, the following scenarios are more favorably indicated for each of the technical applications:}

    \begin{enumerate}[topsep=0pt,itemsep=-1ex,partopsep=1ex,parsep=1ex]
        \item \textit{\textbf{Code Review:} (\ref{sc1}) Descriptive names: Code Review has the highest average rating (4.1) among all four applications. This scenario seems to be particularly suitable for code review activities, as it received the highest average evaluation in this aspect}.

        \item \textit{\textbf{Code Refactoring:} (\ref{sc1}) Descriptive names, (\ref{sc5}) New update, (\ref{sc6}) Showing modifiers, (\ref{sc8}) Risky code, (\ref{sc9}) Problems detected, (\ref{sc10}) Visual reminders receive relatively higher average ratings (ranging from 3.5 to 3.9) across these scenarios. This suggests that participants found these scenarios conducive to code refactoring activities.}
        
        \item \textit{\textbf{Team Efficiency:} (\ref{sc1}) Descriptive names, (\ref{sc8}) Risky code, (\ref{sc9}) Problems detected, (\ref{sc10}) Visual reminders, (\ref{sc11}) Person updated receives relatively higher average ratings (ranging from 3.4 to 3.5) across these scenarios. These scenarios are more conducive to activities related to team collaboration and efficiency.}        
    \end{enumerate}    
\end{quote}

Note that these indications are based on average evaluations provided by participants in the study. The appropriateness of specific scenarios for each application may vary depending on the context and specific objectives of the software development process. Additionally, other factors such as project requirements, team dynamics, and individual preferences should also be taken into account when determining the most suitable scenarios for review, coding, and refactoring activities. To address these questions, inferential statistics were employed using Structural Equation Modeling (SEM), which directly tackles the research question \ref{RQ3}.

\begin{quote}
    \textbf{\ref{RQ3} - What are the rankings of scenario acceptance?}

    \textit{\textbf{\underline{Answer:}} By combining outcomes from various statistical analyses for each hypothesis across different scenarios, we can establish the sequence and categorization of the impact of the scenarios on the utilization of code augmentation. Techniques including Bootstrap and Student's t-test were employed to compute statistical parameters like t-value, p-value, and R-Square ($R^2$), thus facilitating the derivation of rankings and insights, as demonstrated in table \ref{table:ScenarioRaking}, ranked as follows:}
    \begin{enumerate}[topsep=0pt,itemsep=-1ex,partopsep=1ex,parsep=1ex]
        \item \textit{\textbf{Scenario 11} (\ref{sc11}) - Showing the last person that updated some code;}
        \item \textit{\textbf{Scenario 03} (\ref{sc3}) - Variables with bad naming practices;}
        \item \textit{\textbf{Scenario 05} (\ref{sc5}) - New methods and updated methods;}
        \item \textit{\textbf{Scenario 04} (\ref{sc4}) - Recent renames;}
        \item \textit{\textbf{Scenario 06} (\ref{sc6}) - Showing the presence of modifiers in method usage;}
        \item \textit{\textbf{Scenario 01} (\ref{sc1}) - Showing more descriptive names;}
        \item \textit{\textbf{Scenario 07} (\ref{sc7}) - Showing the presence of annotation on method and class usage;}
        \item \textit{\textbf{Scenario 02} (\ref{sc2}) - Showing abbreviated names;}
        \item \textit{\textbf{Scenario 09} (\ref{sc9}) - Problems detected through static analysis tools;}
        \item \textit{\textbf{Scenario 10} (\ref{sc10}) - Visual reminders on API usage.}
    \end{enumerate}   

    \textit{\textbf{Scenario 08} (\ref{sc8} - Calls to risky code) has not undergone validation, thus resulting in its non-classification.}    
\end{quote}

Still based on inferential statistics (SEM), the analysis of various scenarios through statistical methods provides insights into the factors influencing users' intentions to utilize code augmentation. These scenarios are categorized (Section \ref{surveydesignsection}) based on the degree of influence they exert on users' perceptions and decisions. The following explanations elucidate the nature of each scenario and help answer the research question \ref{RQ4}.

\begin{quote}
    \textbf{\ref{RQ4} - What are the key points from each individual scenario?}

    \textit{\textbf{\underline{Answer:}}}

    \begin{enumerate}[topsep=0pt,itemsep=-1ex,partopsep=1ex,parsep=1ex]
        \item \textit{\textbf{Scenario 01 (\ref{sc1}):} Showing more descriptive names;}
        \begin{quote}
            \textit{In Scenario 01 (\ref{sc1}), Hypothesis \ref{h2} (Effort Expectancy $\rightarrow$ Behavioral Intention) is confirmed. However, the insight drawn is that ease of use plays a relatively lower role in influencing users' intentions compared to other scenarios. The willingness of users to use code augmentation is more shaped by their perceived effort required.}    
        \end{quote}
        
        \item \textit{\textbf{Scenario 02 (\ref{sc2}):} Showing abbreviated names;}
        \begin{quote}
            \textit{Similar to Scenario 01 (\ref{sc1}), Hypothesis \ref{h2} (Effort Expectancy $\rightarrow$ Behavioral Intention) is confirmed. Ease of use has a relatively lower influence on users' intentions to adopt code augmentation.}    
        \end{quote}
        
        \item \textit{\textbf{Scenario 03 (\ref{sc3}):} Variables with bad naming practices;}
        \begin{quote}
            \textit{The confirmation of Hypothesis \ref{h6} (Learning Value $\rightarrow$ Behavioral Intention) in this scenario indicates that participants perceive code augmentation as a valuable learning tool. The moderate influence observed suggests that users are inclined to adopt the technology due to its potential to improve their learning experiences.}
        \end{quote}
        
        \item \textit{\textbf{Scenario 04 (\ref{sc4}):} Recent renames;}
        \begin{quote}
            \textit{Hypothesis \ref{h6} (Learning Value $\rightarrow$ Behavioral Intention) is confirmed again in this scenario, reiterating the moderate influence of learning value on users' intentions. Users perceive meaningful learning outcomes as a motivating factor for adopting code augmentation.}    
        \end{quote}        
        
        \item \textit{\textbf{Scenario 05 (\ref{sc5}):} New methods and updated methods;}
        \begin{quote}
            \textit{In this scenario, Hypothesis \ref{h3} (Social Influence $\rightarrow$ Behavioral Intention) is confirmed, indicating that social factors play a moderate role in shaping users' intentions to use code augmentation. The influence of peers and colleagues on users' decisions to adopt the technology is highlighted, showing the significance of social influence.}    
        \end{quote}        
        
        \item \textit{\textbf{Scenario 06 (\ref{sc6}):} Showing the presence of modifiers in method usage;}
        \begin{quote}
            \textit{Two hypotheses, \ref{h2} (Effort Expectancy $\rightarrow$ Behavioral Intention) and \ref{h7} (Habit $\rightarrow$ Behavioral Intention), are confirmed in this scenario. The insight derived suggests that habitual usage, combined with the ease of use, significantly contributes to users' intentions to use code augmentation.}    
        \end{quote}
        
        \item \textit{\textbf{Scenario 07 (\ref{sc7}):} Showing the presence of annotation on method and class usage;}
        \begin{quote}
            \textit{Hypothesis \ref{h2} (Effort Expectancy $\rightarrow$ Behavioral Intention) is again confirmed in Scenario 07 (\ref{sc7}). The insight is that ease of use has a minor impact on users' intentions to use code augmentation in this scenario.}            
        \end{quote}
        
        \item \textit{\textbf{Scenario 08 (\ref{sc8}):} Calls to risky code;}
        \begin{quote}
            \textit{(No hypothesis confirmed)}    
        \end{quote}
                
        \item \textit{\textbf{Scenario 09 (\ref{sc9}):} Problems detected through static analysis tools;}
        \begin{quote}
            \textit{In this scenario, Hypothesis \ref{h2} (Effort Expectancy $\rightarrow$ Behavioral Intention) is confirmed, with a similar insight that perceived ease of use has limited influence on users' intentions to adopt code augmentation.}    
        \end{quote}
        
        \item \textit{\textbf{Scenario 10 (\ref{sc10}):} Visual reminders on API usage;}
        \begin{quote}
            \textit{Hypothesis \ref{h2} (Effort Expectancy $\rightarrow$ Behavioral Intention) is once again confirmed, revealing that ease of use moderately contributes to users' intentions to use code augmentation.}    
        \end{quote}
        
        \item \textit{\textbf{Scenario 11 (\ref{sc11}):} Showing the last person that updated some code;}
        \begin{quote}
            \textit{In this scenario, two key hypotheses, \ref{h1} (Performance Expectancy $\rightarrow$ Behavioral Intention) and \ref{h2} (Effort Expectancy $\rightarrow$ Behavioral Intention), were confirmed. The insight drawn is that participants highly value the performance enhancement and ease of use provided by code augmentation, resulting in a stronger intention to adopt this technology. The positive relationship between performance expectancy and intention reflects users' belief that the technology positively impacts their performance. Likewise, the positive influence of effort expectancy underscores the perceived simplicity and convenience of using code augmentation.}            
        \end{quote}        
    \end{enumerate}

    In summary, the analysis of the results reveals that high-influence scenarios, such as Scenario 11 (\ref{sc11}), play a crucial role in users' intention to adopt code augmentation, driven by positive expectations of performance and ease of use. Additionally, moderate-influence scenarios like Scenarios 03 (\ref{sc3}), 04 (\ref{sc4}), and 05 (\ref{sc5}) also have a significant impact due to the perception of valuable learning and social influence. On the other hand, low-influence scenarios like Scenarios 01 (\ref{sc1}), 02 (\ref{sc2}), 07 (\ref{sc7}), 09 (\ref{sc8}), and 10 (\ref{sc10}) underscore less prominent factors such as ease of use. The unconfirmed hypotheses of \textbf{Facilitating Conditions} and \textbf{Hedonic Motivation} suggest that other elements carry more weight in adoption decisions. Consequently, Scenario 11 (\ref{sc11}) emerges as the most relevant in driving code augmentation adoption based on users' intention to use.
\end{quote}

\subsection{Practical Implications}

Based on the results of the study presented in this paper, we can raise some practical implications of it for different audiences:

\begin{itemize}
    \item \textbf{Educators:} The results of this particular research, especially those evaluated by Scenarios 03 (\ref{sc3}) and 04 (\ref{sc4}) confirming hypothesis \ref{h6}, with t-values indicating statistical significance and R-Squares demonstrating a positive influence of learning value on behavioral intention. Thus, impermanent identifiers can be utilized as an educational resource to augment source code with additional information. This information may draw students' attention to important issues such as code quality, security, and structure. Even demonstrating basic properties of code elements, such as field visibility or whether a method is synchronized, can also prove beneficial when teaching programming.
    
    \item \textbf{API Designers:} The results of this research, particularly through the confirmation of hypotheses \ref{h1}, \ref{h2}, and \ref{h6}, as explicitly expressed in the thematic analysis in statements such as \textit{``help specially new developers''}, suggest that impermanent identifiers can be employed to enhance the usability of APIs by providing additional information when developers are using or maintaining code that utilizes the API. As this application has yet to be explored, various new API design practices and standards may emerge from this opportunity.
    
    \item \textbf{Tool Builders:} This work highlighted the potential and the implementation feasibility of using impermanent identifiers. However, existing tools are not yet mature in this regard, as indicated in the results of the thematic analysis of this study. The thematic analysis identified concerns such as \textit{``Excessive Sensory Stimulation''} and issues with \textit{``Ambiguity''}. Additionally, the API for implementing inlay hints in the IntelliJ IDE, used in codeEmoji, is currently classified as unstable. Our results point to a promising direction for tools that adopt this approach, and tool creators may consider it as an opportunity beyond an alternative or complement to dashboards, other visualizations, and code editor warnings.    
    
    \item \textbf{Developers:} Software developers are the main public for this work, which can directly benefit from an enriched view of the code. As highlighted by the answer to \ref{RQ2}, some scenarios might be more interesting to some activities, like code review or refactoring. When teams start adopting this approach in real projects, some new patterns might be identified in its usage.  
    
    \item \textbf{Researchers:} This work also presents numerous opportunities for researchers interested in exploring the concept of impermanent identifiers in various directions. The following are examples of fields where the proposed approach could be further investigated: Developer eXperience (DX) to address the issues of \textit{``Excessive Sensory Stimulation''} and \textit{``Ambiguity''} identified in thematic analysis, as well as other areas such as code analysis tools, software design, and software engineering education.
\end{itemize} \section{Conclusion}
\label{conclusion}

This article presents an innovative approach to code augmentation using Impermanent identifiers. The goal goes beyond enhancing the software development experience to encompass a comprehensive analysis of the adoption and acceptance of this technology in the development process. For this purpose, this study employed the techniques of the UTAUT model and inferential statistics to assess the acceptance of this approach.

This study's statistical analysis has yielded a thorough comprehension of code augmentation application scenarios and their influence on the software development process. Through evaluating various scenarios and hypotheses, insights have emerged that influence how developers perceive and adopt code augmentation.

The results indicate that various factors with different degrees of influence impact the adoption of code augmentation. In scenarios with a high degree of influence, such as Scenario 11, performance and effort expectations are crucial to the decision to adopt code augmentation. The significance of these factors highlights the necessity of enhancing the functionality and efficacy of code augmentation tools to promote their widespread adoption.

In scenarios with moderate influence, such as Scenarios 03, 04, and 05, factors such as learning value and social influence play a significant role in the intention of developers to employ code augmentation. This emphasizes the significance of fostering collaborative learning and encouraging peer collaboration in order to increase the adoption of code augmentation.

In scenarios with limited influence, such as Scenarios 01, 02, 07, 09, and 10, the impact of usability on the decision to adopt code augmentation is minimal. Although these factors may not be decisive, they still influence the decisions of developers. This underscores the importance of ensuring that code augmentation tools are user-friendly and seamlessly incorporated into the workflow, even when usability is not the determining factor.

The unconfirmed hypotheses regarding facilitating conditions and hedonic motivation suggest that these variables have little effect on adopting code augmentation. This indicates that developers place a greater emphasis on effectiveness, performance, and collaboration than on auxiliary support factors.

Performance expectation, effort expectation, learning value, and social influence emerge as important motivators, whereas facilitating conditions and hedonic motivation exert little or no influence.

The thematic analysis of code augmentation feedback provides insights into its prospective impact on software engineering. The findings emphasize the importance of mitigating risks associated with ambiguity and sensory overload while highlighting the benefits of improved code clarity, legibility, and quality. The recognition of code augmentation's applicability across experience levels and its potential to expedite problem-solving and refactoring bodes well for its incorporation into software development practices. However, this analysis also highlights the need for a deliberate and well-balanced approach that protects the integrity of code comprehension and learning.

Certainly, based on the analysis conducted, the following findings were considered for each research question:

\begin{itemize}
    \item \textbf{Research Question (\ref{RQ1}):} \textit{What is the overall acceptance of code augmentation tools?}
    
    \textbf{Findings:} The study revealed a positive trend in accepting code augmentation technology, with participants showing consensus and favorable reception towards its use. This indicates code augmentation tools' potential value and relevance in software development practices.
\item \textbf{Research Question (\ref{RQ2}):} \textit{Which scenarios are most suitable for Code Review, Refactoring, and Team Efficiency?}
    
    \textbf{Findings:} Specific scenarios were identified as particularly conducive to various technical applications within code augmentation. These findings provide insights for optimizing code review, refactoring, and team efficiency processes, helping developers make informed decisions to improve their development workflows and productivity.
    
\item \textbf{Research Question (\ref{RQ3}):}\textit{ What are the acceptance ratings of the scenarios?}
    
    \textbf{Findings:} Through rigorous statistical analysis, acceptance ratings of the scenarios were established, highlighting their relative impact on the utilization of code augmentation. These ratings offer guidance for prioritizing scenarios for further exploration and development, contributing to the advancement of code augmentation technologies.
    
\item \textbf{Research Question (\ref{RQ4}):} \textit{What are the key points of each individual scenario?}
    
    \textbf{Findings:} Detailed examination of individual scenarios revealed insights into the factors influencing users' intentions to adopt code augmentation. These insights provide information for designing and implementing code augmentation tools that align with users' needs and preferences, paving the way for future research and development efforts aimed at improving the usability and effectiveness of these technologies.    
\end{itemize}

With a strategic approach, code augmentation can play an important role in the software development process. These scenarios provide fundamental guidelines for designing code augmentation tools that meet the requirements of developers, optimize performance, and integrate seamlessly into the workflow.

As software engineering continues to evolve, code augmentation emerges as a modern approach that has the potential to improve the development process, provided that its implementation is meticulously tailored to each project's specific requirements.

Given that the primary focus of this research has been on evaluating the technological acceptance of the tool, and the results indicate a positive acceptance, future work should consider the implementation and practical validation of the approach. Taking into account the initial version of the tool, which encompasses the most fundamental cases of code augmentation, the next stage involves expanding the tool's functionality to encompass more complex scenarios, along with conducting experimental studies aimed at assessing the practical effectiveness of the technique. The analysis of the tangible impact of the technique in various settings emerges as a crucial element in this next phase of the project.

\section*{Acknowledgment of Funding}

The authors would like to extend their profound gratitude to the Free University of Bozen-Bolzano for the generous funding and support provided for the ADVERB project. The collaboration with the institution and the financial support were instrumental in the completion of this study.

This work was partially funded by the ADVERB project, with support from the Free University of Bozen-Bolzano, and we acknowledge the significance of this contribution to the success of this research project.

We appreciate the Free University of Bozen-Bolzano for their belief in our research and their ongoing commitment to advancing knowledge and innovation.
 
\appendix
\section{Survey Analysis Details}
\label{SurveyAnalysis}

The survey was distributed through email by providing a hyperlink to the questionnaire and was accessible for six (6) months to collect responses. Upon the designated period's conclusion, 114 individuals had responded to the survey. 

Surveys like this are very susceptible to motivational bias, which refers to how dedicated the participant is to answering and how well the answers capture the individual's true feelings. To mitigate this occurrence, participants with a standard deviation smaller than 0.5 who did not answer the open questions were deleted from the data set. Six (6) participants were excluded because they met the criteria.

\begin{figure}[!ht]
    \centering
    \resizebox{6.5cm}{!}{
        \begin{tikzpicture}[scale = 1.2]
          \begin{axis}
            [ title  = {},
              ylabel = {Years},
              boxplot/draw direction=y,
              xtick={1,2},
              xticklabels={Age, Experience},
              x tick label style={font=\footnotesize, text width=2.5cm, align=center}
            ]
            \addplot+[mark = *, mark options = {red},
                boxplot prepared={
                  lower whisker     =19, lower quartile    =21, median            =25, average           =28.3, upper quartile    =34.8, upper whisker     =51 }, color = red
            ] coordinates{};
            \addplot+[mark = *,mark options = {blue},
                boxplot prepared={
                  lower whisker     =1, lower quartile    =3, median            =5, average           =8.16, upper quartile    =13, upper whisker     =31 }, color = blue
            ] coordinates{};            
            \end{axis}            
        \end{tikzpicture}
    }
    \caption{Participants’ Age and Experience Box plot}
    \label{fig:boxplotParticipants}
\end{figure}
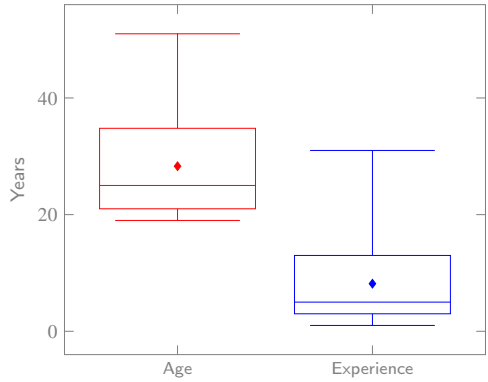

The survey profile reveals a significant predominance of male participants, accounting for 89.8\% (97) of the total sample. On the contrary, the presence of women in the survey is relatively lower, comprising only 8.3\% (9) of the participants. Furthermore, a tiny portion of the sample, equivalent to 0.18\% (2), chose not to respond to the gender-related question. Table~\ref{table:ParticipantsBackground} provides an overview of the subjects' key demographic and experiential characteristics, while Figure~\ref{fig:boxplotParticipants} shows a participants’ Age and Experience Box plot.

\begin{table}[width=.9\linewidth,pos=h]
  \begin{center}		
  \caption{Statistical Overview of Participants' Age and Experience}
  \label{table:ParticipantsBackground}
\begin{tabular}{lccccc}
		  \toprule 
			              & Mean & StDev & Min & Median & Max \\
            \midrule
				 Age        & 28.3 & 8.4 & 19 & 25 & 51 \\
                  Experience & 08.1 & 7.1 & 01 & 05 & 31 \\
		  \bottomrule
    \end{tabular}
    \end{center}
\end{table}

The average age of the participants was 28.3 years, with a standard deviation of 8.4. The age range varied from a minimum of 19 years to a maximum of 51 years. The median age, representing the middle value of the data set, was 25 years. This data distribution indicates a relatively diverse age range among the participants.
    
The average years of experience among the participants was 8.1 years, accompanied by a standard deviation of 7.1. Experience ranged from a minimum of one year to a maximum of 31 years. The median experience was five years. This range highlights a notable variability in the levels of experience within the group of participants.

\subsection{Overview}
\label{survey_results_overview}

Based on the study of the provided situations, a general pattern of consistent scores throughout the numerous scenarios may be detected when Likert's unidimensional scale is used. As already mentioned, participants were asked to rate their level of agreement on a scale ranging from ``Strongly Agree'' (denoted as 5) to ``Strongly Disagree'' (denoted as 1).

According to Table \ref{table:OverviewDescriptive}, the average scores in all situations ranged from 3.1 to 3.6, with a standard deviation between 1.1 and 1.2. This shows that, on average, the situations were evaluated closely, and individual differences in the scores were not excessively large.
In terms of score centrality, the median remained stable at around 3 or 4 in all scenarios. This suggests that, while individual scores varied, the majority of judgments centered on these mean values.

\begin{table}[width=.9\linewidth,pos=h]
  \begin{center}		
  \caption{Descriptive statistics of participant responses across scenarios.}
  \label{table:OverviewDescriptive}
\begin{tabular}{lccccc}
		  \toprule Scenario & Mean & StDev & Min & Median & Max \\
            \midrule
                (\ref{sc1}) Descriptive names & 3.6 & 1.0 & 1 & 4 & 5 \\
                (\ref{sc2}) Abbreviated names & 3.1 & 1.2 & 1 & 3 & 5 \\
                (\ref{sc3}) Bad naming & 3.2 & 1.2 & 1 & 3 & 5 \\
                (\ref{sc4}) Recent renames & 3.1 & 1.1 & 1 & 3 & 5 \\
                (\ref{sc5}) New update & 3.3 & 1.2 & 1 & 3 & 5 \\
                (\ref{sc6}) Showing modifiers & 3.4 & 1.1 & 1 & 4 & 5 \\
                (\ref{sc7}) Showing annotation & 3.3 & 1.1 & 1 & 4 & 5 \\
                (\ref{sc8}) Risky code & 3.5 & 1.1 & 1 & 4 & 5 \\
                (\ref{sc9}) Problems detected & 3.5 & 1.1 & 1 & 4 & 5 \\
                (\ref{sc10}) Visual reminders & 3.5 & 1.1 & 1 & 4 & 5 \\
                (\ref{sc11}) Person updated & 3.1 & 1.2 & 1 & 3 & 5 \\
		  \bottomrule
    \end{tabular}
\end{center}
\end{table}

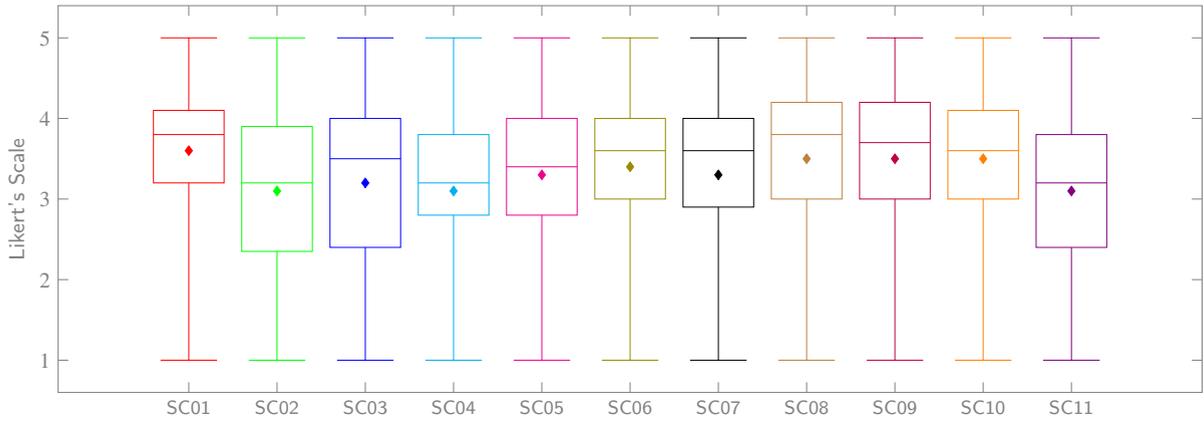
\begin{figure}[h]
    \centering
    \resizebox{16cm}{!}{
        \begin{tikzpicture}[scale = 1.2]
          \begin{axis} 
            [ title  = {},
              ylabel = {Likert’s Scale},
              boxplot/draw direction=y,
              xtick={1,2,3,4,5,6,7,8,9,10,11},
              xticklabels={SC01, SC02, SC03, SC04, SC05, SC06, SC07, SC08, SC09, SC10, SC11},
              x tick label style={font=\footnotesize, text width=2.5cm, align=center},
              cycle list={{red},{green},{blue},{cyan},{magenta},{olive},{black},{brown},{purple},{orange},{violet}},
              x=1.3cm
            ]
            \addplot+[mark = *, mark options = {red},
                boxplot prepared={
                  lower whisker     =1, lower quartile    =3.2, median            =3.8, average           =3.6, upper quartile    =4.1, upper whisker     =5 }, color = red
            ] coordinates{};
            \addplot+[mark = *,mark options = {green},
                boxplot prepared={
                  lower whisker     =1, lower quartile    =2.35, median            =3.2, average           =3.1, upper quartile    =3.9, upper whisker     =5 }, color = green
            ] coordinates{};            
            \addplot+[mark = *,mark options = {blue},
                boxplot prepared={
                  lower whisker     =1, lower quartile    =2.4, median            =3.5, average           =3.2, upper quartile    =4, upper whisker     =5 }, color = blue
            ] coordinates{};
            \addplot+[mark = *,mark options = {cyan},
                boxplot prepared={
                  lower whisker     =1, lower quartile    =2.8, median            =3.2, average           =3.1, upper quartile    =3.8, upper whisker     =5 }, color = cyan
            ] coordinates{};
            \addplot+[mark = *,mark options = {magenta},
                boxplot prepared={
                  lower whisker     =1, lower quartile    =2.8, median            =3.4, average           =3.3, upper quartile    =4, upper whisker     =5 }, color = magenta
            ] coordinates{};
            \addplot+[mark = *,mark options = {olive},
                boxplot prepared={
                  lower whisker     =1, lower quartile    =3, median            =3.6, average           =3.4, upper quartile    =4, upper whisker     =5 }, color = olive
            ] coordinates{};
            \addplot+[mark = *,mark options = {black},
                boxplot prepared={
                  lower whisker     =1, lower quartile    =2.9, median            =3.6, average           =3.3, upper quartile    =4, upper whisker     =5 }, color = black
            ] coordinates{};
            \addplot+[mark = *,mark options = {brown},
                boxplot prepared={
                  lower whisker     =1, lower quartile    =3, median            =3.8, average           =3.5, upper quartile    =4.2, upper whisker     =5 }, color = brown
            ] coordinates{};
            \addplot+[mark = *,mark options = {purple},
                boxplot prepared={
                  lower whisker     =1, lower quartile    =3, median            =3.7, average           =3.5, upper quartile    =4.2, upper whisker     =5 }, color = purple
            ] coordinates{};

            \addplot+[mark = *,mark options = {orange},
                boxplot prepared={
                  lower whisker     =1, lower quartile    =3, median            =3.6, average           =3.5, upper quartile    =4.1, upper whisker     =5 }, color = orange
            ] coordinates{};
            \addplot+[mark = *,mark options = {violet},
                boxplot prepared={
                  lower whisker     =1, lower quartile    =2.4, median            =3.2, average           =3.1, upper quartile    =3.8, upper whisker     =5 }, color = violet
            ] coordinates{};
            \end{axis}            
        \end{tikzpicture}
    }
    \caption{Participants’ Scenarios Box plot}
    \label{fig:boxplotScenarios}
\end{figure}

Figure \ref{fig:boxplotScenarios} shows a scenarios Box plot statistics.

The statistical analysis of the contingency table~\ref{table:OverviewContingency} reveals significant patterns regarding the utilization of code augmentation.

\begin{table}[width=.9\linewidth,pos=h]
  \begin{center}		
  \caption{Scenarios contingency table.}
  \label{table:OverviewContingency}
\resizebox{15cm}{!}{
  \begin{tabular}{lccccc}
		  \toprule Scenario & Strongly Agree & Agree & Don't agree or disagree & Disagree & Strongly disagree \\
                           & (5)            & (4)   & (3)                     & (2) & (1) \\
            \midrule
                (\ref{sc1}) Descriptive names & 408 (23.6\%) & 632 (36.5\%) & 424 (24.5\%) & 207 (11.9\%) & 057 (03.2\%) \\
                (\ref{sc2}) Abbreviated names  & 267 (15.4\%) & 434 (25.1\%) & 459 (26.5\%) & 350 (20.2\%) & 218 (12.6\%)  \\
                (\ref{sc3}) Bad naming  & 294 (17.0\%) & 549 (31.7\%) & 366 (21.1\%) & 349 (20.1\%) & 170 (09.8\%)  \\
                (\ref{sc4}) Recent renames  & 246 (14.2\%) & 430 (24.8\%) & 605 (35.0\%) & 305 (17.6\%) & 142 (08.2\%)  \\
                (\ref{sc5}) New update  & 342 (19.7\%) & 471 (27.2\%) & 469 (27.1\%) & 304 (17.5\%) & 142 (08.2\%)  \\
                (\ref{sc6}) Showing modifiers  & 299 (17.3\%) & 618 (35.7\%) & 459 (26.5\%) & 198 (11.4\%) & 154 (08.9\%)  \\
                (\ref{sc7}) Showing annotation  & 284 (16.4\%) & 595 (34.4\%) & 490 (28.3\%) & 234 (13.5\%) & 125 (07.2\%)  \\
                (\ref{sc8}) Risky code  & 426 (24.6\%) & 552 (31.9\%) & 432 (25.0\%) & 180 (10.4\%) & 138 (07.9\%)  \\
                (\ref{sc9}) Problems detected  & 377 (21.8\%) & 603 (34.8\%) & 442 (25.5\%) & 170 (09.8\%) & 136 (07.8\%)  \\
                (\ref{sc10}) Visual reminders & 370 (21.4\%) & 565 (32.6\%) & 505 (29.2\%) & 147 (08.5\%) & 141 (08.1\%)  \\
                (\ref{sc11}) Person updated & 263 (15.2\%) & 461 (26.6\%) & 453 (26.2\%) & 330 (19.0\%) & 221 (12.7\%)  \\
            \hline
                Total & 3576 (18.8\%) & 5910 (31.0\%) & 5104 (26.8\%) & 2774 (14.5\%) & 1644 (8.6\%)  \\
		  \bottomrule
    \end{tabular}
    }
    \end{center}
\end{table}

\begin{figure}[!ht]
    \centering
    \resizebox{12cm}{!}{
        \begin{tikzpicture}[scale = 1.2]
            \begin{axis}[
                ybar stacked,
            	bar width=15pt,
            	nodes near coords,
                    nodes near coords align={anchor=center},every node near coord/.append style={font=\footnotesize},
                totals/.style={nodes near coords={}},
                enlargelimits=0.15,
                legend style={at={(0.5,-0.20)},
                  anchor=north,legend columns=-1},
                ylabel={\# Participants},
                symbolic x coords={SC01, SC02, SC03, SC04, SC05, SC06, SC07, SC08, SC09, SC10, SC11},
                xtick=data,
                x tick label style={rotate=45,anchor=east},
                x=1cm
                ]
            \addplot+[ybar] plot coordinates {(SC01,408) (SC02,267) (SC03,294) (SC04,246) (SC05,342) (SC06,299) (SC07,284) (SC08,426) (SC09,377) (SC10,370) (SC11,263)};
            \addplot+[ybar] plot coordinates {(SC01,632) (SC02,434) (SC03,549) (SC04,430) (SC05,471) (SC06,618) (SC07,595) (SC08,552) (SC09,603) (SC10,565) (SC11,461)};
            \addplot+[ybar] plot coordinates {(SC01,424) (SC02,459) (SC03,366) (SC04,605) (SC05,469) (SC06,459) (SC07,490) (SC08,432) (SC09,442) (SC10,505) (SC11,453)};
            \addplot+[ybar] plot coordinates {(SC01,207) (SC02,350) (SC03,349) (SC04,305) (SC05,304) (SC06,198) (SC07,234) (SC08,180) (SC09,170) (SC10,147) (SC11,330)};
            \addplot+[ybar,brown!20!black,fill=brown!80!white] plot coordinates {(SC01,057) (SC02,218) (SC03,170) (SC04,142) (SC05,142) (SC06,154) (SC07,125) (SC08,138) (SC09,136) (SC10,141) (SC11,221)};
            \legend{\strut Strongly Agree, \strut Agree, \strut  Don't agree or disagree, \strut Disagree, \strut Strongly disagree}
            \end{axis}

        \end{tikzpicture}
    }
    \caption{Graphical scenarios contingency table}
    \label{fig:barchartScenarioContingency}
\end{figure}
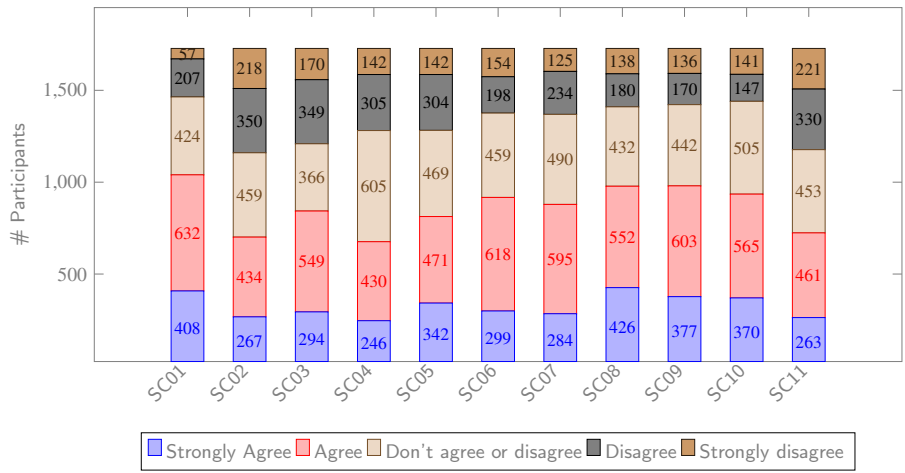

There is a clear tendency toward high levels of agreement across all scenarios. For example, in Scenario 01, the majority of responses are divided between ``Strongly Agree'' (23.6\%) and ``Agree'' (36.5\%), showing a high level of agreement among participants. Similarly, in other scenarios (02–10), the ``Strongly Agree'' and ``Agree'' categories predominate, indicating an overall tendency toward agreement. Figure~\ref{fig:barchartScenarioContingency} depicts a Graphical Scenarios Contingency table.

However, this trend is not consistent. In Scenario 11, for example, the responses are more evenly distributed, spanning all categories in equivalent quantities. This indicates a more significant difference of opinion in this specific circumstance.

In some circumstances, the ``Don't Agree nor Disagree'' category stands out, reflecting nuances in participants' perspectives and maybe a lack of clarity in their positions. Similarly, the inclusion of the ``Strongly Disagree'' category in variable proportions across different scenarios emphasizes the existence of opposing opinions, particularly in specific contexts.

The constancy in median categories and generally consistent standard deviations across scenarios suggest that the opinions of the participants are stable. On the other hand, fluctuation in the percentage distribution of each category across multiple scenarios demonstrates the complexities of individual perceptions in the face of diverse given contexts.

In summary, despite the occurrence of disagreement in some instances, the data show a consistent pattern of moderate to strong agreement. This underscores the importance of considering nuances in interpretations while acknowledging the overall degree of agreement among participants.

According to Table~\ref{table:TechnicalAppDesc}, the ``New Code'' application has a moderate average score of 3.3, a relatively constant standard deviation of 1.1, and a median of 3. The ``Code Review'' application, on the other hand, has a slightly higher average score of 3.5, as well as a moderate dispersion indicated by a standard deviation of 1.2 and a higher median of 4. Similarly, the ``Code Refactoring'' program scores 3.4 on average, with a corresponding standard deviation of 1.2 and a median of 4. Finally, with an average score of 3.3, a steady standard deviation of 1.1, and a median of 3, the ``Team Efficiency'' application follows the pattern. 

\begin{table}[width=.9\linewidth,pos=h]
  \begin{center}	
\caption{Descriptive statistics of participant responses for different technical applications.}
  \label{table:TechnicalAppDesc}
\begin{tabular}{lccccc}
		  \toprule & Mean & StDev & Min & Median & Max \\
            \midrule
                (NC) New Code         & 3.3 & 1.1 & 1 & 3 & 5 \\
                (CR) Code Review      & 3.5 & 1.2 & 1 & 4 & 5 \\
                (CF) Code Refactoring & 3.4 & 1.2 & 1 & 4 & 5 \\
                (TE) Team Efficiency  & 3.3 & 1.1 & 1 & 3 & 5 \\
		  \bottomrule
    \end{tabular}
    \end{center}
\end{table}

\begin{figure}[!ht]
    \centering
    \resizebox{7cm}{!}{
        \begin{tikzpicture}[scale = 1.2]
          \begin{axis} 
            [ title  = {},
              ylabel = {Likert’s Scale},
              boxplot/draw direction=y,
              xtick={1,2,3,4},
              xticklabels={NC, CR, CF, TE},
              x tick label style={font=\footnotesize, text width=2.5cm, align=center},
              cycle list={{red},{green},{blue},{orange}}
]
            \addplot+[mark = *, mark options = {red},
                boxplot prepared={
                  lower whisker     =1, lower quartile    =3, median            =3.3, average           =3.3, upper quartile    =3.9, upper whisker     =5 }, color = red
            ] coordinates{};
            \addplot+[mark = *,mark options = {green},
                boxplot prepared={
                  lower whisker     =1, lower quartile    =3.1, median            =3.6, average           =3.5, upper quartile    =4.1, upper whisker     =5 }, color = green
            ] coordinates{};            
            \addplot+[mark = *,mark options = {blue},
                boxplot prepared={
                  lower whisker     =1, lower quartile    =3.1, median            =3.5, average           =3.4, upper quartile    =4, upper whisker     =5 }, color = blue
            ] coordinates{};
            \addplot+[mark = *,mark options = {orange},
                boxplot prepared={
                  lower whisker     =1, lower quartile    =3, median            =3.4, average           =3.3, upper quartile    =3.9, upper whisker     =5 }, color = orange
            ] coordinates{};            
            \end{axis}            
        \end{tikzpicture}
    }
    \caption{Technical applications Box plot}
    \label{fig:boxplotTechnical}
\end{figure}
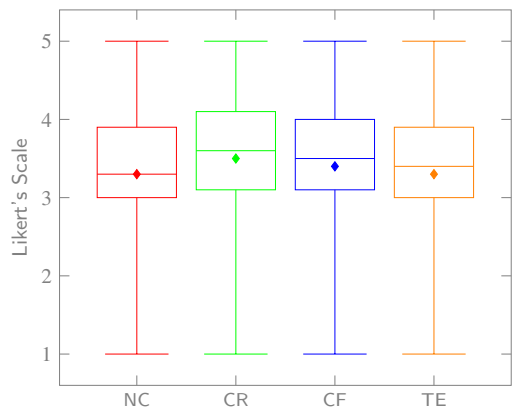

Figure \ref{fig:boxplotTechnical} shows a Technical applications Box plot statistics.

These results collectively indicate a moderate level of agreement across the board, with the ``Code Review'' and ``Code Refactoring'' applications exhibiting slightly higher mean scores and medians, potentially suggesting a stronger positive perception among participants compared to the ``New Code'' and ``Team Efficiency'' applications. The consistent standard deviations imply relatively stable responses within each application. This analysis emphasizes the importance of considering these nuances while interpreting the participants' perspectives on various technical applications.

Table~\ref{table:TechnicalApp} provides a summary of the average evaluations of the technical applications in each of the 11 scenarios analyzed. Each scenario reflects the averages assigned to the four applications: ``New Code'', ``Code Review'', ``Code Refactoring'', and ``Team Efficiency''. The averages were calculated based on participants' responses, where they indicated their perceptions of the different applications in each scenario.

\begin{table}[width=.9\linewidth,pos=h]
  \begin{center}		
\caption{Average evaluations of technical applications across analyzed scenarios.}  
  \label{table:TechnicalApp}
\begin{tabular}{lcccc}
		  \toprule Scenario & New Code & Code Review & Code Refactoring & Team Efficiency \\ 
            \midrule
                (\ref{sc1}) Descriptive names  & 3.4 & 4.1 & 3.9 & 3.5 \\
                (\ref{sc2}) Abbreviated names  & 3.2 & 3.1 & 3.1 & 3.0 \\
                (\ref{sc3}) Bad naming  & 3.1 & 3.3 & 3.3 & 3.1 \\
                (\ref{sc4}) Recent renames  & 3.0 & 3.2 & 3.3 & 3.1 \\
                (\ref{sc5}) New update  & 3.2 & 3.5 & 3.5 & 3.3 \\
                (\ref{sc6}) Showing modifiers  & 3.3 & 3.6 & 3.5 & 3.3 \\
                (\ref{sc7}) Showing annotation  & 3.3 & 3.5 & 3.4 & 3.3 \\
                (\ref{sc8}) Risky code  & 3.5 & 3.5 & 3.5 & 3.5 \\
                (\ref{sc9}) Problems detected  & 3.5 & 3.6 & 3.6 & 3.4 \\
                (\ref{sc10}) Visual reminders & 3.5 & 3.6 & 3.5 & 3.4 \\
                (\ref{sc11}) Person updated & 2.9 & 3.2 & 3.0 & 3.4 \\
		  \bottomrule
    \end{tabular}
\end{center}
\end{table}

The analysis reveals variations in the averages in scenarios and applications. For example, in Scenario 1, the ``New Code'' application received an average of 3.4, while ``Code Review'', ``Code Refactoring'', and ``Team Efficiency'' received averages of 4.1, 3.9, and 3.5, respectively. However, in Scenario 11, the averages were generally lower, with ``New Code'' averaging 2.9, ``Code Review'' at 3.2, ``Code Refactoring'' at 3.0, and ``Team Efficiency'' at 3.4.

\begin{figure}[!ht]
    \centering
    \resizebox{12cm}{!}{
        \begin{tikzpicture}[scale = 1.2]
            \begin{axis}[
                ybar stacked,
            	bar width=15pt,
            	nodes near coords,
                    nodes near coords align={anchor=center},every node near coord/.append style={font=\footnotesize},
                totals/.style={nodes near coords={}},
                enlargelimits=0.15,
                legend style={at={(0.5,-0.20)},
                  anchor=north,legend columns=-1},
                ylabel={\# Participants},
                symbolic x coords={SC01, SC02, SC03, SC04, SC05, SC06, SC07, SC08, SC09, SC10, SC11},
                xtick=data,
                x tick label style={rotate=45,anchor=east},
                x=1cm
                ]
            \addplot+[ybar] plot coordinates {(SC01,3.4) (SC02,3.2) (SC03,3.1) (SC04,3.0) (SC05,3.2) (SC06,3.3) (SC07,3.3) (SC08,3.5) (SC09,3.5) (SC10,3.5) (SC11,2.9)};
            \addplot+[ybar] plot coordinates {(SC01,4.1) (SC02,3.1) (SC03,3.3) (SC04,3.2) (SC05,3.5) (SC06,3.6) (SC07,3.5) (SC08,3.5) (SC09,3.6) (SC10,3.6) (SC11,3.2)};
            \addplot+[ybar] plot coordinates {(SC01,3.9) (SC02,3.1) (SC03,3.3) (SC04,3.3) (SC05,3.5) (SC06,3.5) (SC07,3.4) (SC08,3.5) (SC09,3.6) (SC10,3.5) (SC11,3.0)};
            \addplot+[ybar] plot coordinates {(SC01,3.5) (SC02,3.0) (SC03,3.1) (SC04,3.1) (SC05,3.3) (SC06,3.3) (SC07,3.3) (SC08,3.5) (SC09,3.4) (SC10,3.4) (SC11,3.4)};
            
            \legend{\strut New Code, \strut Code Review, \strut  Code Refactoring, \strut Team Efficiency}
            \end{axis}

        \end{tikzpicture}
    }
    \caption{Graphical Scenarios Contingency table}
    \label{fig:barchartThecnical}
\end{figure}
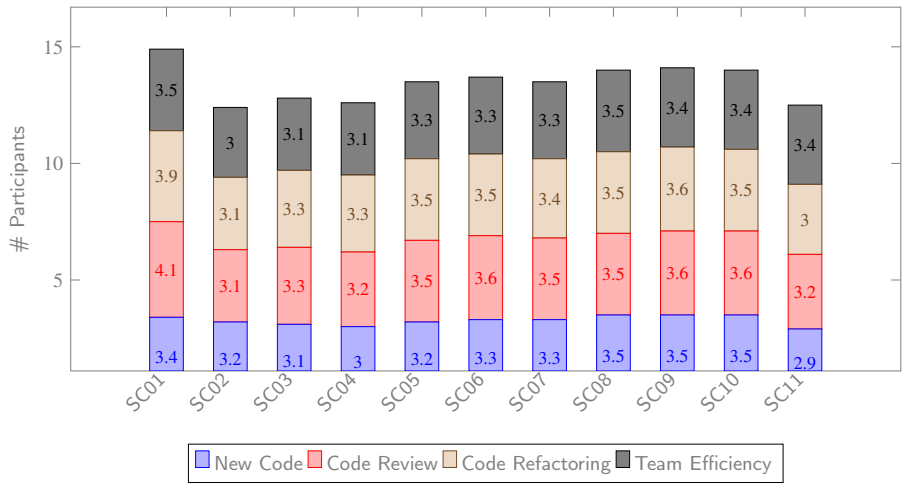

The figure \ref{fig:barchartThecnical} shows a technical applications across analyzed scenarios.

To better understand individual perceptions and the diversity of opinions in each scenario, and with the aim of analyzing the most relevant features in the intention to use code augmentation, the study was conducted through SEM.

\subsection{Structural Equation Modeling - SEM}

Structural Equation Modeling (SEM) is an advanced statistical approach used to investigate complex interactions between observed and unobserved variables within a model \citep{English2006}. It is frequently used in research domains such as social sciences, psychology, education, and economics to test and validate complex theories or hypotheses \citep{English2006}.

SEM permits the evaluation of both direct and indirect causal relationships between variables and the fit of the model to the observed data. It is useful for comprehending the interrelationships between variables, identifying latent factors (unobserved) that can influence observed variables, and evaluating the magnitude and direction of their associations \citep{Kaplan2009}.

Based on the UTAUT 2 model presented in Section~\ref{UTAUT}, the analysis was constructed for each of the 11 scenarios. The gender variable has been removed from the model because the research data show that 89.8\% of the participants are male. In other words, there isn't enough data to apply the Gender moderator. The statistical rationale behind not using the Gender moderator is based on statistical power and sample size concepts. For a moderator variable to be properly examined and have a meaningful impact on the model, the sample size for each moderator category must be sufficiently broad and diverse \cite{Kristopher2007}. 

The model evaluation was carried out by examining Cronbach's Alpha (CA) for internal consistency, followed by Composite Reliability (CR) as a measure of internal consistency. Convergent validity was assessed through Average Variance Extracted (AVE), and discriminant validity was estimated using the cross-loadings criterion \citep{Henseler2009}.

This approach facilitated an examination of individual reliability, delving into the correlations among the questionnaire items. The extent of their correlation reflects the explanatory strength of a particular variable. Expanding on this, it is worth noting that a higher correlation signifies a more robust relationship between variables, indicating that changes in one variable are more likely to result in corresponding changes in another. This insight provides a deeper understanding of how interconnected the variables within the model are and underscores the importance of considering the interplay between them when interpreting the results. Furthermore, a strong correlation suggests that the measured constructs are consistent and aligned, enhancing the overall reliability and validity of the analysis.

According to Table~\ref{table:ReliabilityTesting}, it is evident that the AVE tests with values $\geq 0.5$ \citep{fornell1981} and Cronbach's Alpha $\geq 0.6$ \citep{hill2002} yielded values within the appropriate standards. Similar to Cronbach's Alpha, Composite Reliability is a measure of reliability. This coefficient also achieved results that exceeded the acceptable threshold of $\geq 0.7$ \citep{chin1998} for all constructs.

\begin{table}[width=.9\linewidth,pos=h]
  \begin{center}		
\caption{Reliability measures for various constructs across scenarios.}
  \label{table:ReliabilityTesting}
\resizebox{16cm}{!}{
  \begin{tabular}{lc*{7}{c}}
		  \toprule \multirow{2}{*}{Scenario} 
                        &  & \textbf{Performance} & \textbf{Effort}     & \textbf{Social}      & \textbf{Facilitanting} & \textbf{Hedonic}    & \textbf{Learning} & \multirow{2}{*}{Habit} \\
                        &  & \textbf{Expectancy}  & \textbf{Expectancy} & \textbf{Incluence}   & \textbf{Conditions}    & \textbf{Motivation} & \textbf{Value}    & \\
                        &  & (\ref{CT-PerfExp})   & (\ref{CT-ExpEff})   & (\ref{CT-SocialInf}) & (\ref{CT-FaciCond})    & (\ref{CT-Hedonic})  & (\ref{CT-LeanVal}) & (\ref{CT-Habit}) \\
            \midrule
                \multirow{3}{*}{(\ref{sc1}) Descriptive names}
                    & CA   & 0.801 & 0.692 & 0.683 & 0.769 & 0.888 & 0.727 & 0.570 \\
                    & CR   & 0.909 & 0.866 & 0.860 & 0.896 & 0.946 & 0.869 & 0.823 \\
                    & AVE & 0.834 & 0.764 & 0.755 & 0.811  & 0.898 & 0.770 & 0.699 \\
            \hline
                \multirow{3}{*}{(\ref{sc2}) Abbreviated names}
                    & CA   & 0.906 & N/D & 0.819 & 0.893 & 0.932 & 0.885 & 0.695 \\
                    & CR   & 0.942 & N/D & 0.917 & 0.949 & 0.967 & 0.946 & 0.866 \\
                    & AVE & 0.843 & N/D & 0.846 & 0.904  & 0.936 & 0.897 & 0.764 \\
            \hline
                \multirow{3}{*}{(\ref{sc3}) Bad naming}
                    & CA   & 0.868 & 0.817 & 0.863 & 0.941 & 0.875 & 0.928 & 0.842 \\
                    & CR   & 0.938 & 0.916 & 0.936 & 0.971 & 0.941 & 0.965 & 0.927 \\
                    & AVE & 0.883 & 0.845 & 0.880 & 0.944  & 0.888 & 0.933 & 0.864 \\
            \hline
                \multirow{3}{*}{(\ref{sc4}) Recent renames} 
                    & CA   & 0.856 & 0.862 & 0.835 & 0.891 & 0.937 & 0.952 & 0.895 \\
                    & CR   & 0.931 & 0.935 & 0.923 & 0.948 & 0.970 & 0.976 & 0.950 \\
                    & AVE & 0.872 & 0.879 & 0.858 & 0.901  & 0.941 & 0.954 & 0.905 \\
            \hline
                \multirow{3}{*}{(\ref{sc5}) New update} 
                    & CA   & 0.936 & 0.893 & 0.903 & 0.914 & 0.939 & 0.969 & 0.921 \\
                    & CR   & 0.969 & 0.949 & 0.953 & 0.959 & 0.970 & 0.985 & 0.962 \\
                    & AVE & 0.940 & 0.903 & 0.911 & 0.920  & 0.942 & 0.970 & 0.927 \\
            \hline
                \multirow{3}{*}{(\ref{sc6}) Showing modifiers}
                    & CA   & 0.909 & 0.943 & 0.868 & 0.956 & 0.940 & 0.905 & 0.831 \\
                    & CR   & 0.956 & 0.964 & 0.938 & 0.978 & 0.971 & 0.954 & 0.922 \\
                    & AVE & 0.916 & 0.898 & 0.883 & 0.957  & 0.944 & 0.913 & 0.855 \\
            \hline
                \multirow{3}{*}{(\ref{sc7}) Showing annotation} 
                    & CA   & N/D & 0.940 & 0.880 & 0.957 & 0.911 & 0.938 & 0.888 \\
                    & CR   & N/D & 0.962 & 0.943 & 0.979 & 0.958 & 0.970 & 0.946 \\
                    & AVE & N/D & 0.894 & 0.893 & 0.959  & 0.919 & 0.942 & 0.898 \\
            \hline
                \multirow{3}{*}{(\ref{sc8}) Risky code} 
                    & CA   & 0.943 & 0.951 & 0.917 & 0.922 & 0.943 & 0.905 & 0.916 \\
                    & CR   & 0.964 & 0.968 & 0.960 & 0.962 & 0.972 & 0.954 & 0.960 \\
                    & AVE & 0.898 & 0.910 & 0.923 & 0.927  & 0.946 & 0.913 & 0.922 \\
            \hline
                \multirow{3}{*}{(\ref{sc9}) Problems detected} 
                    & CA  & 0.963 & 0.963 & 0.904 & 0.909 & 0.940 & 0.914 & 0.891 \\
                    & CR  & 0.982 & 0.976 & 0.954 & 0.956 & 0.971 & 0.958 & 0.948 \\
                    & AVE & 0.964 & 0.932 & 0.913 & 0.916 & 0.943 & 0.920 & 0.901 \\
            \hline
                \multirow{3}{*}{(\ref{sc10}) Visual reminders}
                    & CA  & 0.928 & 0.917 & 0.909 & 0.913 & 0.933 & 0.917 & 0.809 \\
                    & CR  & 0.965 & 0.960 & 0.956 & 0.958 & 0.967 & 0.960 & 0.912 \\
                    & AVE & 0.933 & 0.923 & 0.917 & 0.919  & 0.937 & 0.923 & 0.839 \\
            \hline
                \multirow{3}{*}{(\ref{sc11}) Person updated}
                    & CA   & 0.929 & 0.936 & 0.891 & 0.947 & 0.929 & 0.959 & 0.857 \\
                    & CR   & 0.955 & 0.959 & 0.948 & 0.974 & 0.966 & 0.980 & 0.933 \\
                    & AVE & 0.876 & 0.887 & 0.902 & 0.950  & 0.934 & 0.960 & 0.875 \\            
		  \bottomrule
    \end{tabular}
    }
    \end{center}
\end{table}

The consistency and high reliability observed in the CR and AVE. values across various scenarios strengthen the confidence in the study's measurements. The consistently strong reliability measures across constructs and scenarios indicate that the survey items measure the intended constructs with high consistency and accuracy.

Discriminant validity is a pivotal concept in the analysis of constructs within research. By adhering to the \citet{fornell1981}, a common standard for discriminant validity, the research ensures that the measures employed to gauge various constructs are indeed separate and not highly correlated. This reinforces the reliability and validity of the interpretations drawn from the conducted analyses.

In addition, the discriminant validity test assesses indicators to verify whether they differ from each other. Both the estimated correlation matrix and the square root of the average extracted variance from constructs are evaluated. As mentioned, the discriminant validity data obtained in this study adhere to \citet{fornell1981}, in which the value in each column is greater than those below it, representing the correlation between the constructs \citep{hair2014, Ringle2014}.

Table~\ref{table:Scen1DisValidity} presents the Estimated Correlation Matrix for Scenario 01. Within this context, discriminant validity is assessed through two primary approaches: the estimated correlation matrix and the square root of the average variance extracted (AVE) from constructs.

\begin{table}[width=.9\linewidth,pos=h]
  \begin{center}		
  \caption{Scenario 1: Discriminant validity metrics.}
  \label{table:Scen1DisValidity}
\begin{tabular}{l*{9}{c}}
		  \toprule & \textbf{BI*} & \ref{CT-ExpEff} & \ref{CT-FaciCond} & \ref{CT-Habit} & \ref{CT-Hedonic} & \ref{CT-LeanVal} & \ref{CT-PerfExp} & \ref{CT-SocialInf} \\
            \midrule
                \textbf{(BI*) Behavioral Intention}                     & \textbf{1.000} &  &  &  &  &  &  &  \\
                \textbf{(\ref{CT-ExpEff}) Effort Expectancy}            & 0.484 & \textbf{0.874} &  &  &  &  &  &  \\
                \textbf{(\ref{CT-FaciCond}) Facilitanting Conditions}   & 0.334 & 0.612 & \textbf{0.901} &  &  &  &  &  \\
                \textbf{(\ref{CT-Habit}) Habit}                         & 0.334 & 0.505 & 0.266 & \textbf{0.836} &  &  &  &  \\
                \textbf{(\ref{CT-Hedonic}) Hedonic Motivation}          & 0.334 & 0.580 & 0.538 & 0.691 & \textbf{0.948} &  &  &  \\
                \textbf{(\ref{CT-LeanVal}) Learning Value}              & 0.270 & 0.477 & 0.492 & 0.503 & 0.574 & \textbf{0.877} &  &  \\
                \textbf{(\ref{CT-PerfExp}) Performance Expectancy}      & 0.402 & 0.812 & 0.561 & 0.582 & 0.638 & 0.476 & \textbf{0.913} &  \\
                \textbf{(\ref{CT-SocialInf}) Social Incluence}          & 0.370 & 0.623 & 0.498 & 0.547 & 0.629 & 0.494 & 0.595 & \textbf{0.869} \\ 
		  \bottomrule
            \multicolumn{9}{l}{\textit{* - Behavioral Intention}}
    \end{tabular}
\end{center}
\end{table}

In the case of the Estimated Correlation Matrix, correlation coefficients among constructs are examined. Discriminant validity is confirmed when the correlation coefficients among constructs are less than 1, indicating that they are not perfectly correlated. This suggests that the measures are distinct enough to represent different characteristics. Similarly, for the AVE, discriminant validity is met when the square root of the AVE for each construct is greater than the correlations between that construct and the others. This means that the shared variance among constructs is smaller than the variance captured by each construct individually, validating their distinctiveness.

By verifying whether the data meet the \citet{fornell1981} criterion, a common standard for discriminant validity, the research ensures that the measures used to assess different constructs are genuinely separate and not highly correlated. This reinforces the reliability and validity of interpretations drawn from the conducted analyses.

For each of the 11 scenarios, we conducted a discriminant validity analysis. To meet the criteria suggested by \citet{fornell1981}, which requires that each value in a column be greater than the values below it, we performed cross-load analyses to eliminate predictor variables if necessary.

An analysis was performed using the bootstrapping technique to investigate the values of the Student's t-test \cite{STUDENT1908, field2013} for each of the 11 scenarios. This investigation sought to determine the correlations between the constructs reflecting the hypotheses explored in this study. Bootstrapping is a statistical approach that involves creating several samples from the original data to perform a more accurate and reliable analysis \citep{hair2014}. According to \citet{hair2014}, this adaptable method helps to calculate crucial statistics, build confidence intervals, and rigorously test hypotheses without relying on tight assumptions about data distribution.
In this paper, reference values for hypotheses accepted using the Student's t-test were t-value $\geq\ $ 1.96 and p-value < 0.05, respectively \citep{hair2014}. The bootstrapping analysis is presented in Table~\ref{table:Bootstrapping}, which highlights the scenarios that were able to confirm the defined hypotheses.

\begin{table}[width=.9\linewidth,pos=h]
\vspace{-2mm}
  \begin{center}	
\caption{Results of bootstrapping analysis for hypotheses across scenarios}
  \label{table:Bootstrapping}
\resizebox{16cm}{!}{
  \begin{tabular}{lcl*{3}{c}}
		  \toprule & \multirow{2}{*}{\textbf{Hypothesis}} & \textbf{Confirmed}   & \multirow{2}{*}{\textbf{t-values}} & \textbf{p-values} & \multirow{2}{*}{\textbf{$R^2$}} \\
                  &                                      & \textbf{by scenario} &          & (Position) &          \\
            \midrule
                Performance Expectancy $\rightarrow$ BI* & \ref{h1} & (\ref{sc11}) Person updated & 3.467 & 0.001 (01) & 0.652 \\
            \hline
                \multirow{7}{*}{Effort Expectancy $\rightarrow$ BI*}
                                                       & \multirow{7}{*}{\ref{h2}} 
                                                            & (\ref{sc1}) Descriptive names & 2.705 & 0.007 (04) & 0.251 \\
                                                       &    & (\ref{sc2}) Abbreviated names & 2.442 & 0.015 (05) & 0.547 \\
                                                       &    & (\ref{sc6}) Showing modifiers & 2.010 & 0.045 (11) & 0.600 \\
                                                       &    & (\ref{sc7}) Showing annotation & 2.090 & 0.037 (09) & 0.546 \\
                                                       &    & (\ref{sc9}) Problems detected & 2.052 & 0.040 (10) & 0.584 \\
                                                       &    & (\ref{sc10}) Visual reminders & 2.707 & 0.007 (03) & 0.612 \\
                                                       &    & (\ref{sc11}) Person updated & 2.111 & 0.035 (08) & 0.652 \\
            \hline
                Social Incluence $\rightarrow$ BI*         & \ref{h3} & (\ref{sc5}) New update & 2.009 & 0.045 (12) & 0.622 \\
            \hline
                Facilitanting Conditions $\rightarrow$ BI* & \ref{h4} & \textbf{\textit{Not confirmed}} &  &  &  \\
            \hline
                Hedonic Motivation $\rightarrow$ BI*       & \ref{h5} & \textbf{\textit{Not confirmed}} &  &  &  \\
            \hline
                Learning Value $\rightarrow$ BI*           & \ref{h6} & (\ref{sc3}) Bad naming & 2.723 & 0.006 (02) & 0.738 \\
                                                         &    & (\ref{sc4}) Recent renames & 2.318 & 0.020 (07) & 0.616 \\                
            \hline
                Habit $\rightarrow$ BI*                    & \ref{h7} & (\ref{sc6}) Showing modifiers & 2.430 & 0.015 (06) & 0.600 \\
            \hline
                \multicolumn{6}{l}{\textbf{MODERATORS}} \\
            \hline
                Age x Facilitanting Conditions $\rightarrow$ BI*        & \textbf{\ref{h4a}$A_{1}$} & \textbf{\textit{Not confirmed}} &  &  &  \\
                Age x Learning Value $\rightarrow$ BI*                  & \textbf{\ref{h5a}$A_{1}$} & \textbf{\textit{Not confirmed}} &  &  &  \\
                Age x Hedonic Motivation $\rightarrow$ BI*              & \textbf{\ref{h6a}$A_{1}$} & \textbf{\textit{Not confirmed}} &  &  &  \\
                Age x Habit $\rightarrow$ BI*                           & \textbf{\ref{h7a}$A_{1}$} & \textbf{\textit{Not confirmed}} &  &  &  \\
                Experiency x Facilitanting Conditions $\rightarrow$ BI* & \textbf{\ref{h4e}$E_{1}$} & \textbf{\textit{Not confirmed}} &  &  &  \\
                Experiency x Learning Value $\rightarrow$ BI*           & \textbf{\ref{h5e}$E_{1}$} & \textbf{\textit{Not confirmed}} &  &  &  \\
                Experiency x Hedonic Motivation $\rightarrow$ BI*       & \textbf{\ref{h6e}$E_{1}$} & \textbf{\textit{Not confirmed}} &  &  &  \\
                Experiency x Habit $\rightarrow$ BI*                    & \textbf{\ref{h7e}$E_{1}$} & \textbf{\textit{Not confirmed}} &  &  &  \\
		  \bottomrule
            \textit{* - Behavioral Intention} & & & & \multicolumn{1}{l}{$R^2$ Mean}   & 0.585 \\
                                              & & & & \multicolumn{1}{l}{$R^2$ Median} & 0.606 \\
                                              & & & & \multicolumn{1}{l}{$R^2$ StDev}  & 0.117 \\
    \end{tabular}
    }
    \end{center}
\end{table}
 
\bibliographystyle{cas-model2-names}
\bibliography{cas-refs}

\end{document}